\documentclass[useAMS,usenatbib]{mn2e}
\usepackage{ctable}
\usepackage{float}
\usepackage{graphicx}
\usepackage{color}
\usepackage{amsmath}
\usepackage{ctable}
\newcommand{\kms}{\ensuremath{\mathrm{km\,s^{-1}}}}

\usepackage{graphicx}
\usepackage{amsmath}
\begin{document}
\title[A non-degenerate set of optical IMF indicators]{The stellar IMF in early-type galaxies \\ from a non-degenerate set of optical line indices.\\ } 
\author[Spiniello et al.]{Chiara Spiniello$^{1, 2}$\thanks{E-mail:
spiniello@astro.rug.nl }, Scott Trager$^{1}$, L\'{e}on V.E. Koopmans$^{1}$, Charlie Conroy$^{2}$ \\
\\
$^{1}$Kapteyn Astronomical Institute, University of Groningen, PO Box 800,9700 AV Groningen, the Netherlands\\
$^{2}$Now at: Max-Planck Institute for Astrophysics, Karl-Schwarzschild-Strasse 1, 85740 Garching, Germany\\
$^{3}$Department of Astronomy \& Astrophysics, University of California, Santa Cruz, CA, USA}
\date{Accepted Year Month Day. Received Year Month Day; in original form Year Month Day}
\pagerange{\pageref{firstpage}--\pageref{lastpage}} \pubyear{2002}
\maketitle

\label{firstpage}

\begin{abstract}
We investigate the optical spectral region of spectra of $\sim 1000$
stars searching for IMF-sensitive features to constrain the
low-mass end of the initial mass function (IMF) slope in elliptical
galaxies.  The use of indicators bluer than NIR features (NaI, CaT,
Wing-Ford FeH) is crucial if we want to compare our observations to
optical simple stellar population (SSP) models. 
We use the MILES stellar library (\citealt{SanchezBlazquez2006}) 
in the wavelength range 3500--7500 \AA\, to select
indices that are sensitive to cool dwarf stars and that do not or only
weakly depend on age and metallicity.  We find several promising
indices of molecular TiO and CaH lines. 
In this wavelength range, the response of a change in the effective
temperature of the cool red giant (RGB) population is similar to the
response of a change in the number of dwarf stars in the galaxy. We
therefore investigate the degeneracy between IMF variation and $\Delta {\rm
T_{\rm eff, RGB}}$ and show that it is possible to break this
degeneracy with the new IMF indicators defined here.
In particular, we define a CaH1 index around $\lambda$6380 \AA\, that 
arises purely from cool dwarfs, does not strongly depend on age 
and is anti-correlated with [$\alpha$/Fe]. 
This index allows the determination of the low-mass end of the IMF slope from 
integrated-light measurements when combined with different TiO lines 
and age- and metallicity-dependent features such as H$\beta$, Mg$b$, Fe5270 and Fe5335. %and CaT.
The use of several indicators is crucial to break degeneracies between IMF variations, 
age, abundance pattern and effective temperature of the cool red giant (RGB) population.
We measure line-index strengths of our new optical IMF
indicators in the \citet{Conroy2012} SSP models and compare these with index strengths
of the same spectral features in a sample of stacked Sloan Digital Sky
Survey (SDSS) early-type galaxy (ETG) spectra with varying velocity dispersions.  
Using different indicators, we find a clear trend of a steepening IMF with increasing
velocity dispersion from 150 to $310\,\mathrm{km}\,\mathrm{s}^{-1}$ described by the linear
equation $x = (2.3\pm 0.1) \, \log\,\sigma_{200} + (2.13\pm 0.15)$,
where $x$ is the IMF slope and $\sigma_{200}$ is the central stellar
velocity dispersion measured in units of $200\,\mathrm{km}\,\mathrm{s}^{-1}$. 
We test the robustness of this relation by repeating the analysis with ten 
different sets of indicators. We found that the NaD feature has the largest impact on the IMF slope, 
if we assume solar [Na/Fe] abundance. 
By including NaD the slope of the linear relation increases by 0.3 ($2.6 \pm 0.2$). 
We compute the ``IMF mismatch" parameter as the ratio of stellar mass-to-light 
ratio predicted from the $x-\sigma_{200}$ relation to that inferred from SSP models 
assuming a Salpeter IMF and find good agreement with independent published results. 
%{\bf Finally, we compute the $({\rm M/L})_{\star}$--$\sigma_{\star}$ relation 
%in the $\sigma$-range [200-350] $\kms$, finding 
%$\log (\alpha)_{\rm IMF} = (1.05\pm 0.2) \, \log\,\sigma_{\star} - (2.5\pm 0.4)$.}
\end{abstract}


\section{Introduction}

The stellar contribution to the mass budgets of early-type galaxies
(ETGs) is a crucial ingredient to fully understand the internal
structure, the formation and the evolution of these massive galaxies
(e.g.\ \citealt{Blumenthal1984}). In ETGs the stellar mass-to-light
ratio ($\Upsilon_{\star}$) scales with the luminous mass of the system 
(\citealt{Grillo2009, Barnabe2011, Dutton2012, Cappellari2012, Tortora2013}), 
but it has also been shown recently that, for the
assumption of a universal initial mass function (IMF), the dark matter
(DM) fraction in the internal region of these systems increases with
the mass of the galaxy (e.g.\ \citealt{Zaritsky2006, Auger2010, Treu2010, Barnabe2011}).
Disentangling the relative contributions of baryonic and dark matter
constituents of ETGs is therefore crucial to fully comprehend the
processes that shape the hierarchical galaxy formation scenario (e.g.\ \citealt{White1978, Davis1985, Frenk1985}). 

A quantitative study of the luminous unresolved stellar content of distant galaxies 
requires the use of stellar population synthesis
models (\citealt{Worthey1994, Renzini2006, Conroy2013}). 
By comparing colors, line-strength indices or full spectral energy distributions
(SEDs) in galaxy spectra with predictions from single stellar
population (SSP) models (which assume a single epoch of star formation
rather than an extended SFH), it is possible to derive stellar
parameters such as luminosity-weighted age, metallicity, $\Upsilon_{\star}$,
IMF slope and elemental abundances for a galaxy (\citealt{Worthey1994, Trager2000, Trager2000b}).  
However, it remains difficult to trace a
galaxy's full star formation history (SFH), to break the
age-metallicity degeneracy (\citealt{Worthey1994}) and to study
possible correlation of IMF slope and $\alpha-$enhancement variation
with galaxy masses (\citealt{Spinrad1962, Spinrad1971, Cohen1978, Frogel1978, Frogel1980,
Faber1980, Carter1986, Hardy1988, Couture1993, Worthey1994, Cenarro2003}). 

In the last two decades a crucial assumption has been
made in constraining the star formation history of galaxies through
SSPs: the IMF is assumed to be universal and equal to that of our
Milky Way (\citealt{Kroupa2001, Chabrier2003, Bastian2010}). 
In the past years, evidence has emerged that the IMF might evolve
 (\citealt{Dave2008, vanDokkum2008}) and may depend on the (stellar)
mass of the system  (e.g.\ \citealt{Treu2010, Auger2010b, 
Napolitano2010, vandokkum2010, Spiniello2011, Spiniello2012}). 
Recently, \citet{vandokkum2010}, hereafter vDC10, have suggested that low-mass stars ($\le 0.3\,M_{\odot}$) 
could be much more prevalent in massive early-type galaxies (ETGs) than previously thought. 
This could imply that the increase in the $\Upsilon_{\star}$ with galaxy mass is due to a steeper low-mass 
end of the IMF rather than an increasing fraction of internal DM 
(\citealt{Treu2010, Auger2010, Barnabe2011, Dutton2012, Cappellari2012, Spiniello2012}).
 
Strong absorption features that vary with surface gravity at fixed effective temperature 
and that can be used to count stars with masses $\leq 0.3\,M_{\odot}$  (M dwarfs) 
are mainly present in the red-optical and near-infrared spectral region (vDC10, \citealt{Smith2012}).
At these wavelengths galaxy stellar emission is dominated by evolved stellar populations, 
i.e. red giant branch (RGB) and asymptotic giant branch (AGB) stars (e.g.\ \citealt{Worthey1994, Renzini2006}), 
whose physics is not yet well understood, because their lifetime is very
short and mass-losses are very high (e.g.\ \citealt{Reimers1975}). 
 Moreover, AGB stars are also highly variable  (e.g.\ \citealt{Blocker1995}). 
 In the optical spectral region, on the other hand, M dwarfs contribute at most 5\% to the total
light of the integrated spectrum, despite dominating the total
stellar mass budget in galaxies (\citealt{Worthey1994}). The light in ETGs is
dominated by K and M giants, that, to the first order, have spectra
similar to the one of an M dwarf, because of the similar spectral
type.  However the spectra of M dwarfs and M giants with similar
effective temperature show minor but important differences.  Careful
line-index strength measurements of multiple spectral indices (or full
spectral fitting: \citealt{Conroy2012b}) are necessary to reveal
these spectral differences at the percent level and to break
degeneracies, enabling one to observe variations of IMF with galaxy
mass if present.

Two new SSP models have recently been published specifically for the
purpose of measuring the IMF slope down to $\sim 0.1\, {M}_{\odot}$ for
old, metal-rich stellar populations.  \citet{Conroy2012}, hereafter CvD12, 
built models over the wavelength interval $0.35\, \mu m \,< \lambda < 2.4\,\mu m$ at a resolving power of R $\sim 2000$,
using a combination of two empirical stellar libraries (MILES,
\citealt{SanchezBlazquez2006}; and IRTF, \citealt{Cushing2005})
and three sets of isochrones.
\citet{Vazdekis2012}  recently built the MIUSCAT models, an
extension in wavelength of the \citet{Vazdekis2003,Vazdekis2003b, Vazdekis2010} models. 

All other available models in the NIR are based on theoretical
atmospheres and stellar libraries (\citealt{Maraston2005, Bruzual2003}), 
which have only been tested by fitting broadband colors and do not
reproduce line indices measurements of clusters and galaxies
(\citealt{Lyubenova2012}). The CvD12 and MIUSCAT models are therefore the
state-of-the-art of SSP models.

Both models allow for different IMF slopes and a range of ages, but
while the CvD12 models use solar metallicity isochrones and synthesize
models with different abundance patterns, MIUSCAT models use
different total metallicities but do not allow one to change the
abundances which are fixed to solar.  
Non-solar abundance patterns appear to be a necessary ingredient to properly 
assess the non-universality of the IMF, especially in massive ETGs, 
which may have undergone a star formation histories 
different from the solar-neighbourhood (\citealt{Peterson1976,Peletier1989,Worthey1992, Trager2000b, Arrigoni2010}). 
The CvD12 models allow for a variation in [$\alpha$/Fe] 
as well as the abundance pattern of 11 different elements. 
This is particularly important in the case of 
sodium lines, especially when including the NaD index. 
As shown in \citet{Conroy2012,Conroy2012b} and confirmed in \citet{Spiniello2012}, 
more massive systems seem to be Na-enhanced ([Na/Fe] $\sim 0.3$--$0.4$ dex), 
consistent with values obtained for the bulge of the Milky Way (on average [Na/Fe] $\sim 0.2$, \citealt{Fulbright2007}); 
see Spiniello et al., in prep., for a detailed study of the impact of models with and without 
variable abundance patterns on the determination of the low-mass IMF slope in ETGs. 
We therefore use the CvD12 models in this study to have the required ability to decouple 
IMF variations from abundance variations. 
Moreover, CvD12 also include the possibility to change the effective temperature of the red giant branch
(RGB), taking into account that the isochrones should change with abundance pattern.

In the blue region (3500--7500 \AA) the
CvD12 and MIUSCAT models use the same empirical spectral library (MILES), while in
the red and NIR they make use of two different libraries (CVD12 uses IRTF, 
MIUSCAT uses Indo-US, \citealt{Valdes2004}, and CaT, \citealt{Cenarro2001}).  
The use of optical indicators allows us to reduce the uncertainties 
caused by different assumptions of different SSP models in the synthesis process (e.g.\ their spectral libraries).
%This is one of the main reasons why it is important to look at the blue region of the models.
%It's important to caution, perhaps at the end of this paragraph, that the bluer the feature the more sensitive you are to *earlier* spectral types, and so you are not directy probing the bottom of the IMF.  That is the tradeoff with going bluer.

\citet{Conroy2012b} clearly show that there is information on
the IMF slope in the blue spectral range, the goal of this work is to
recover and quantify this information.  We find a set of optical
IMF-sensitive spectral indicators that allow us to decouple the
effects of a varying IMF from age, metallicity and/or elemental
abundance variations, and the effective temperature of the RGB, when
studying the stellar populations of massive ETGs.
A direct comparison of the variation of index strengths with IMF
slope predicted from the SSP models with the variation of index strengths
determined from SDSS spectra allows us to understand whether the
recent suggestion that the IMF steepens with increasing galaxy velocity
dispersion is genuine or arises from a misunderstanding of the main
ingredients of SSP models.

In this work (i) we focus on line-index measurements rather than full
spectral fitting, which avoids issues with spectral calibration when
comparing to observations that might have been poorly calibrated and
(ii) we assume an SSP, rather than an extended SFH, which is still a
strong assumption and will be properly addressed in future works.  The
paper is organized as follows. In Section~2 we present a new set of
IMF-sensitive indicators along with a brief introduction of the
MILES library.  In Section~3 we compare line-index variations as
function of the IMF slope from single MILES stars and the CvD12 SSP
model.  In Section~4 we compare the models with SDSS galaxies. We
summarize our findings and discuss our conclusions in Section~5.

\section[NewIndices]{New optical IMF-sensitive indices}

Stellar spectral features that show different strengths in M-dwarfs
and cool giants, including NaI, Wing-Ford FeH, CaT and CaI, can
potentially reveal a galaxy's low-mass stellar content in spectra of
unresolved stellar populations (e.g.\ \citealt{Spinrad1962, Faber1980, Schiavon1997,Schiavon1997b,Schiavon2000, Cenarro2003}).  
These are mainly present in the red-optical and near-infrared part of the
spectra.  However, optical spectra are easier to obtain with 
current spectrographs than NIR (1--$2\,\mu\mathrm{m}$) spectra.  Moreover,
near-infrared spectra are contaminated by the presence of strong sky lines and
telluric absorption arising from water vapour (e.g.\ \citealt{Stevenson1994}). 
A proper measurement of equivalent widths (EW) in the red part of the spectrum relies heavily 
on the correct removal of these lines.
It is therefore important to search for IMF-sensitive features in
the blue-optical spectral region, where the SSP models are less
affected by these uncertainties and different assumptions.
However one has to keep in mind that blue-optical spectral features
are more sensitive to stars of earlier spectral types.
%and  therefore by
%using {\sl only} blue indicators we are less sensitive to the very
%low-mass end of the IMF than studies using red/NIR spectra.

\subsection{Searching for IMF-sensitive features in the MILES Library}

We searched for spectral features that would be sensitive to IMF variations in SSPs 
by searching for features that appear solely in cool dwarf stars in the MILES empirical spectral
stellar library (\citealt{SanchezBlazquez2006}), the same
library used by the CvD12 and MIUSCAT models. The MILES library
consists of $\sim1000$ stellar spectra obtained at the 2.5m INT
telescope.  It covers the wavelength range 3525--7500~\AA\, with
$2.51$~\AA\, (FWHM) spectral resolution and spans a large range in
atmospheric parameters: [Fe/H]= [$-2.86$, 1.65], $\log{g}$=[$-0.2$, 5.5]
and ${T_{\rm eff}}$= [2747, 36000].
Using these spectra, we identified a number of potentially interesting
stellar features that could be used to constrain the low-mass end of
the IMF. We plotted the ratio between spectra of a cool giant (${T_{\rm eff}} \sim 3300K$) 
and a cool dwarf (${T_{\rm eff}} \sim 2800K$), 
both with roughly solar metallicity, to select these spectral 
features. We inspected this ratio looking for wavelength regions with large residuals. 
We identified seven interesting 
regions, including four new and previously unidentified features, and 
defined absorption-line indices for these.  Moreover, we ran a 
principal component analysis on the SSP models to isolate features that 
strongly depend on temperature and gravity.  The final IMF-sensitive indices we 
have found and use here are listed in Table~\ref{tab:definition}. 
We note that TiO and CaH indices are measured in magnitudes, while NaD is measured in \AA.

\begin{figure*}
\centering
\includegraphics[height=20cm]{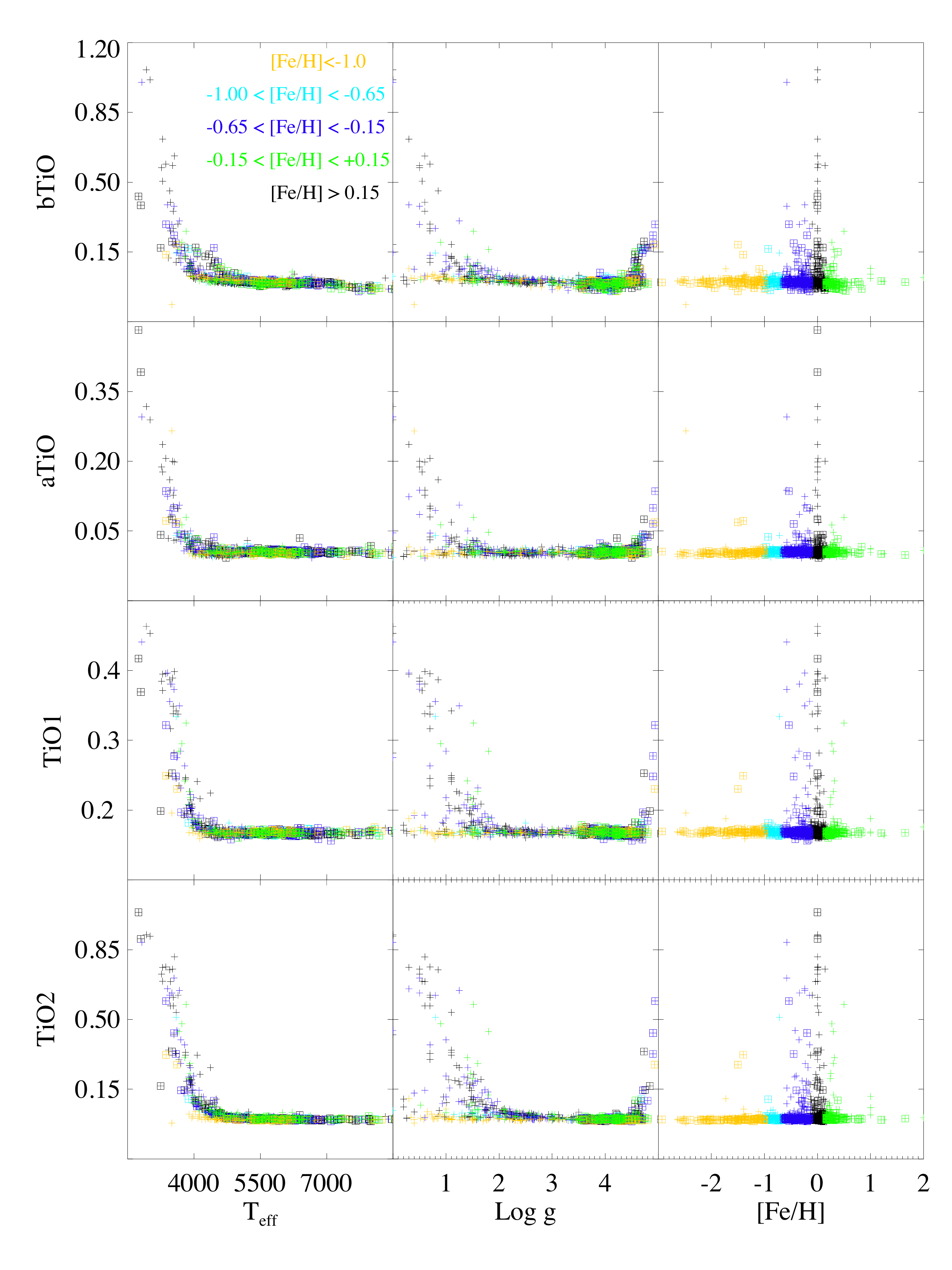}
\caption{Index strengths of TiO indices for single stars from the MILES
  library plotted as function of temperature, gravity, and [Fe/H], at the original MILES resolution.
  The points are coded by $\log{g}$: crosses are mostly giants, and 
  squares are dwarfs.  The colors refer to [Fe/H], as seen in the
  upper-left panel. All TiO indices are very strong in cool stars 
  ($T < 4100 K$). From the middle panel, it is clear that the
  indices are stronger in giants (i.e.\ $\log{g} < 3.0$) and M dwarfs
  (i.e.\ $\log{g} > 3.8$) than in warmer main-sequence stars.}
\label{fig:miles1}
\end{figure*}

\begin{figure*}
\centering
\includegraphics[height=16cm]{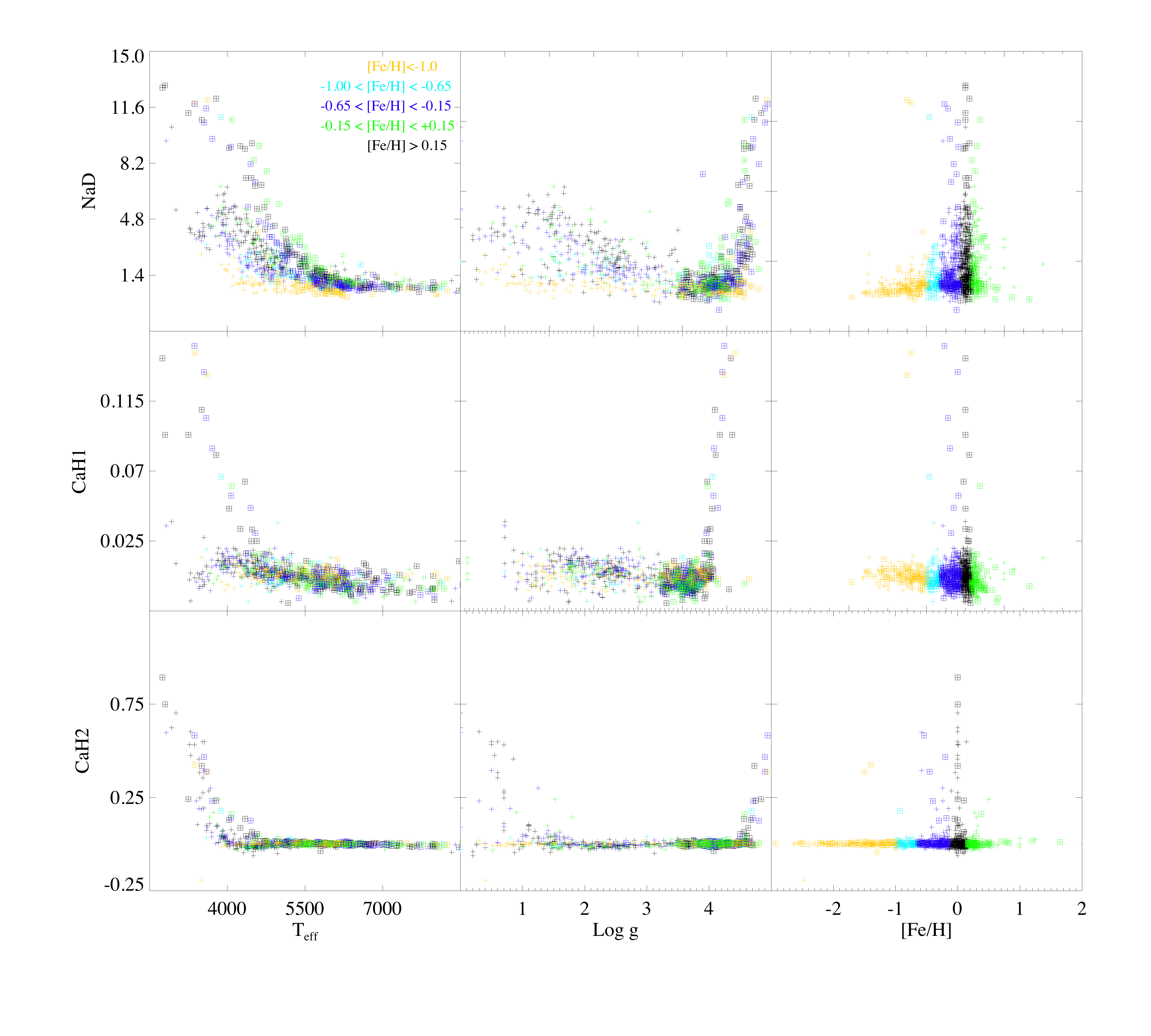}
\caption{ 
EWs of CaH and NaD indices for single stars from the MILES
  library plotted as function of temperature, gravity, and [Fe/H], at the original MILES resolution. 
  The points are coded by $\log{g}$: crosses are mostly giants, and 
  squares are dwarfs.  The colors refer to [Fe/H], as seen in the
  upper-left panel. 
The middle row shows the only feature which 
is very strong only in cool stars ($T < 4500 K$) with high gravity (i.e.\ $\log{g} > 3.8$): CaH1. 
This index allows us to count the M-dwarf population in a galaxy spectrum.}  
\label{fig:miles2}
\end{figure*}

Figures~\ref{fig:miles1} and ~\ref{fig:miles2} show index
strengths of these indices for single MILES stars as a
function of temperature (left panels), gravity (middle panels) and
metallicity (right panels). The figures clearly show that all the selected
stellar absorption features are very weak in main-sequence stars (MS)
and intermediate-temperature stars and are strong in cool dwarfs and
giants -- except CaH1, which is only strong in cool dwarfs--.

More than half of the selected indices come from TiO molecular
absorption bands (Fig.~\ref{fig:miles1}).  The effect of TiO lines in
stellar atmospheres has been extensively studied over the years
(\citealt{Jorgensen1994, Allard2000, Valenti2010}) and red
TiO lines have already been recognized to be good IMF indicators
 (\citealt{Spiniello2012}, Chen et al. 2013, in prep.).  
Our bluest feature is around $\lambda \sim 4770$~\AA: we call it blue-TiO and
define a new index (bTiO, Table~\ref{tab:definition}). 
Although a similar index was already used in \citet{Serven2005}, who named it Mg4780, 
we are confident that this absorption is due to TiO molecular lines, whose strength is very large in very cool dwarf and giant stars 
and smaller for intermediate-mass M stars, as shown in the middle panel of Figure~\ref{fig:miles1}.  
The aTiO index, defined
around the $\alpha$-TiO feature (\citealt{Jorgensen1994}), is also
strongly IMF-sensitive.  
A possible issue with this
  feature is the presence in its central bandpass of the very strong
  [OI] $\lambda5577$~\AA\ sky emission line. Any contamination from
  sky lines could affect the response to gravity of the spectral index
  defined around aTiO.  However from Figure~\ref{fig:miles1} it
  appears that the subtraction of the emission line in the library stars is
  sufficient, as the behaviour of aTiO seems to be similar to that of
  bTiO and TiO2 (i.e., stronger EWs for cool giants and dwarfs).  
  The TiO2 feature ($\lambda \sim6230$ \AA) was already found to be
  promising in constraining the low-mass end of the IMF slope  (see \citealt{Spiniello2012}), 
  but it also shows a weak dependence on
  metallicity and $\alpha$-element variation (Chen et al.\ 2013, in
  prep).  
  The NaD feature, although gravity-sensitive, also depends
  strongly on abundance  (in particular this index is 
  $\sim 4$ times more sensitive to variation in the [Na/Fe] 
  abundance than to variation of the IMF slope, as shown 
  in Spiniello et al. (2013, in prep.) and may be contaminated by the interstellar medium (e.g.\ \citealt{Spiniello2012}). 

\begin{table}
\caption{Definition of the IMF-sensitive indices used in this paper. 
TiO and CaH indices are measured in magnitudes, while NaD is measured in \AA\ } 
\centering 
\begin{tabular}{lcc} 
\hline
\hline 
{  \bf Index   }  & {  \bf  Central band (\AA)}& { \bf   Pseudo-continua (\AA)}\\
[0.5ex]	% inserts table %heading \hline 1&50&837&970 \\ 2&47&877&230 \\ 3&31&25 &415 \\ 4 & 35 & 144 & 2356 \\ 5 & 45 & 300 & 556 \\ [1ex] 
\hline 
bTiO		& 		4758.500   -- 4800.000 	&  4742.750  -- 4756.500    \\
				&&  4827.875  -- 4847.875\\
aTiO		&		5445.000   --  5600.000	& 5420.000   --  5442.000	\\
				&& 5630.000    -- 5655.000\\

NaD		&		5876.875    --  5909.375 &  5860.625    --  5875.625 \\
			&&	 5922.125    -- 5948.125 \\   
TiO1	  &		5936.625    -- 5994.125 &  5816.625    -- 5875.625  \\
	&&		6038.625    -- 6103.625 \\
TiO2 		&	  6189.625    -- 6272.125 &	 6066.625    -- 6141.625 \\
		&& 6372.625    -- 6415.125 \\
CaH1  &  6357.500    -- 6401.750  	&	6342.125    -- 6356.500  \\
&&	6408.500    -- 6429.750 \\
CaH2  &  6775.000    -- 6900.000  	&	6510.000    -- 6539.250  \\
&& 7017.000    -- 7064.000 \\
\hline
\hline 
\end{tabular} 
\label{tab:definition}
\end{table}

\begin{figure*}
\centering
\includegraphics[height=20cm]{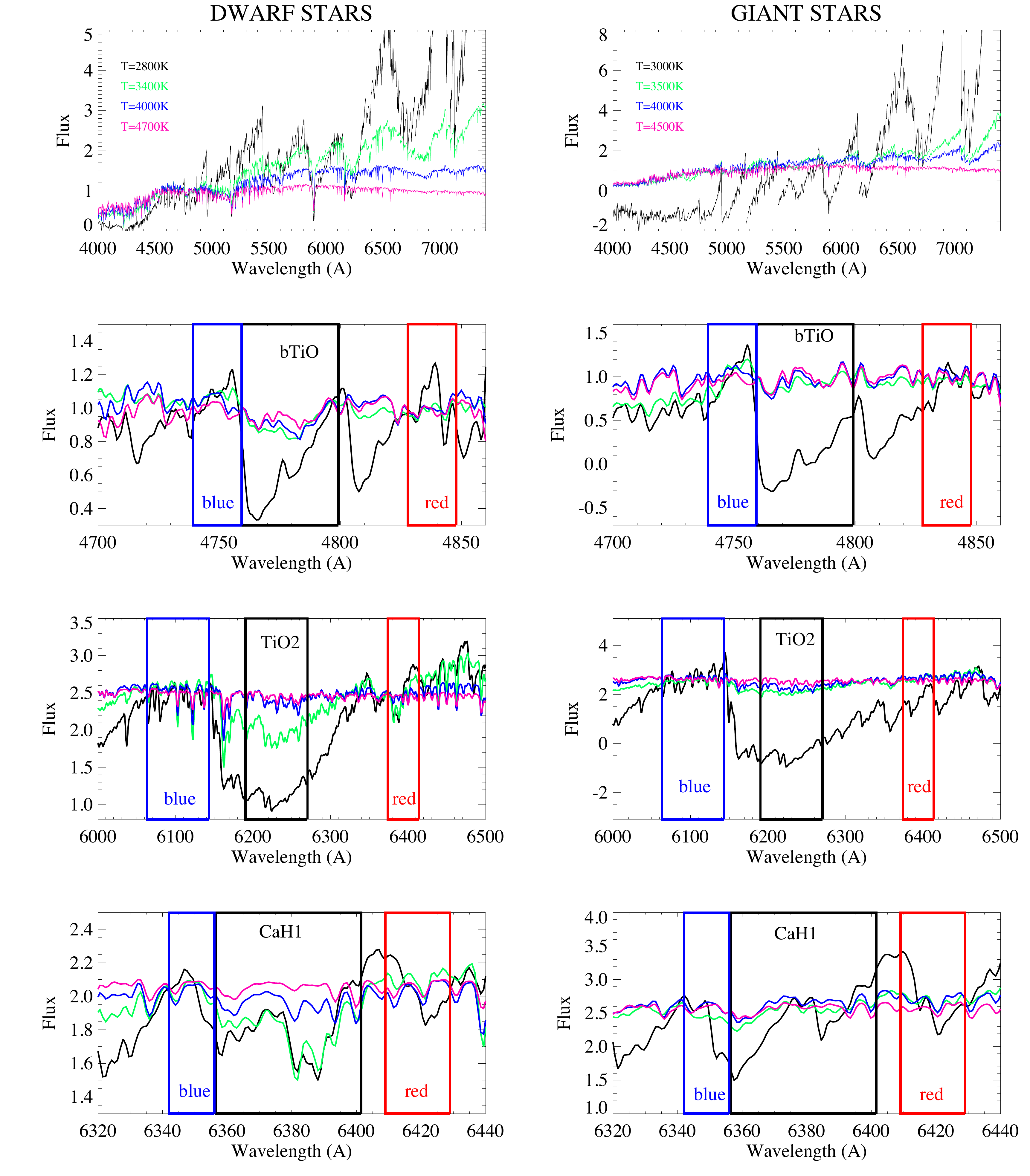}
\caption{{\sl Left panels:} Individual MILES dwarf stars with similar
  surface gravities ($\log{g} \sim 5$) and different $T_{\rm
    eff}$. {\sl Right panels:} Individual MILES giants with similar
  surface gravities ($\log{g} \sim 0.5$) and different $T_{\rm eff}$.
  In the bottom three rows, we show a zoom into different absorption
  feature regions. Red and blue boxes are the blue and red
  pseudo-continua bands of the indices. Both bTiO and TiO2 absorption
  are stronger in cool stars but are present in both dwarfs and
  giants. The CaH1 index instead only comes from cool dwarfs. }
\label{fig:Teff_dwarfs}
\end{figure*}

As a sanity check, we show spectra of single dwarf stars with similar
gravity ($\log{g} = 4.6$--$5$) at different values of ${T_{\rm eff}}$ in the left
panels of Figure~\ref{fig:Teff_dwarfs} in order to ensure that these
indices are stronger in cool dwarfs than in warm ones. The right
panels of Figure~\ref{fig:Teff_dwarfs} show instead spectra of single
cool giants with a range of effective temperatures.  bTiO and TiO2 are
clearly stronger in cool stars but they behave similarly in giants and
dwarfs, therefore they do not allow one to disentangle between these
two classes of stars and between a top-heavy and a bottom-heavy IMF by
themselves.  \\
\indent However, the CaH1 index around $\lambda$6380 \AA\,
(Fig.~\ref{fig:miles2}), is strong only in cool dwarfs. We believe that this feature is the 
most robust optical feature to estimate the IMF in this particular wavelength region, because it 
allows us to disentangle the contribution of RGB stars from the
cool M-dwarf population when combined with age- and metallicity-sensitive features. 
CaH features were first detected in spectra of M dwarfs by \citet{Fowler1907} 
and then studied in detail by \citet{Ohman1934}.
Already in that paper \"{O}hman pointed out that the CaH band at 6390$\,$\AA$\,$(corresponding 
to our CaH1 index) is strong in M-dwarf spectra and almost absent in M giants.
\citet{Mould1976} realized that the behaviour of CaH in stellar atmospheres is strongly influenced 
by gas pressure over the formation of the molecule, and therefore he used the CaH band as 
a luminosity indicator of cool stars in low resolution spectra.
He also studied the sensitivity of the CaH band strength to gravity, clearly confirming 
that CaH bands are very weak or absent in giants. 
Finally, \citet{Barbuy1993} presented a study of the intensity of CaH bands as a 
function of stellar temperature and gravity. Their results are qualitatively consistent 
with ours in the sense that CaH absorption lines are stronger in dwarfs and tend to disappear in giants. 
We refer to section 5.4 of \citet{Conroy2012} for a useful sketch of the basic physics involved.

This brief historical overview demonstrates that it has been known for a hundred years that 
certain spectral features depend strongly on surface gravity at fixed effective temperature and therefore  
betray the presence of faint M dwarfs in integrated light spectra.

%Although we call this index CaH, we suspect that the gravity sensitivity may come again from TiO absorption.
%In the redder part of the optical spectra, we found a very promising CaH absorbtion band 
%(CaH1 $\lambda 6380$ \AA) which seems to be very strong only in cool dwarfs stars.

\begin{figure}
\centering
\includegraphics[height=5cm]{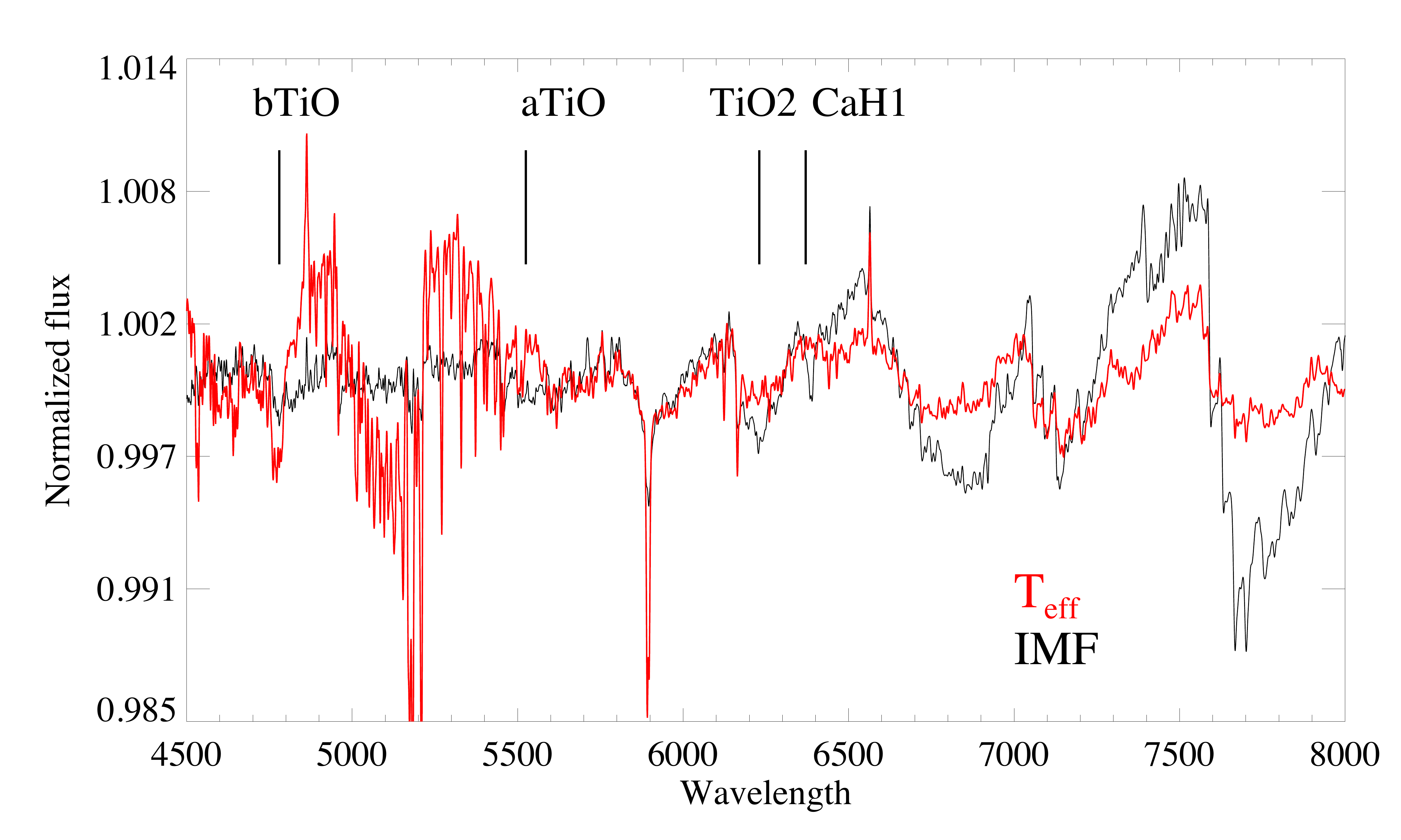}
\caption{Relative effect on the spectrum of IMF and $T_{\rm eff}$ variations. Blueward of $\sim7500$ \AA\, the effects 
are almost perfectly degenerate while they decouple redward of TiO2 and CaH1. 
The response of the two curves however is different around specific features. 
This makes their combination a powerful tool to decouple the variations.}
\label{fig:Teff}
\end{figure}

From Figure~\ref{fig:Teff_dwarfs}, it is clear that the spectrum of a
cool giant is similar to that of a cool dwarf.  We
therefore also investigate the response of variation in the effective
temperature of red-giant branch versus the response in variation of
the IMF slope, to ensure that the degeneracy between ${T_{ \rm eff,
  RGB}}$ and IMF slope can be broken by our optical IMF indicators.
Indeed, one of the problems that one faces when using indices bluer that $\sim7500$ \AA\, 
is that the effects of a varying IMF and of a varying $T_{\rm eff}$ of the isochrones 
are very similar, as shown in Figure~\ref{fig:Teff}. 

 %. 
In the figure we plot the continuum normalized response to IMF variation and variation of 
the value of $T_{\rm eff}$ of the RGBs as a function of wavelength.
As expected, the IMF variation is \emph{almost} completely overwhelmed by $T_{\rm eff}$ variation.
%luckily, redward of $\sim7500$, $T_{\rm eff}$ and IMF effects become very different. 
However, the two curves differ in their responses around bTiO, aTiO and CaH1.  
Therefore only the use of the full set of lines allows one to separate IMF and $T_{\rm eff}$ effects.
By combining blue spectral features that strongly depend on age, metallicity and elemental abundance 
with surface-gravity-sensitive features, such as TiO and CaH lines, we can jointly 
constrain these effects and the low-mass end of the IMF 
in integrated optical-light spectra of distant ETGs.

\section{SSP models}
\begin{figure*}

\includegraphics[height=15cm]{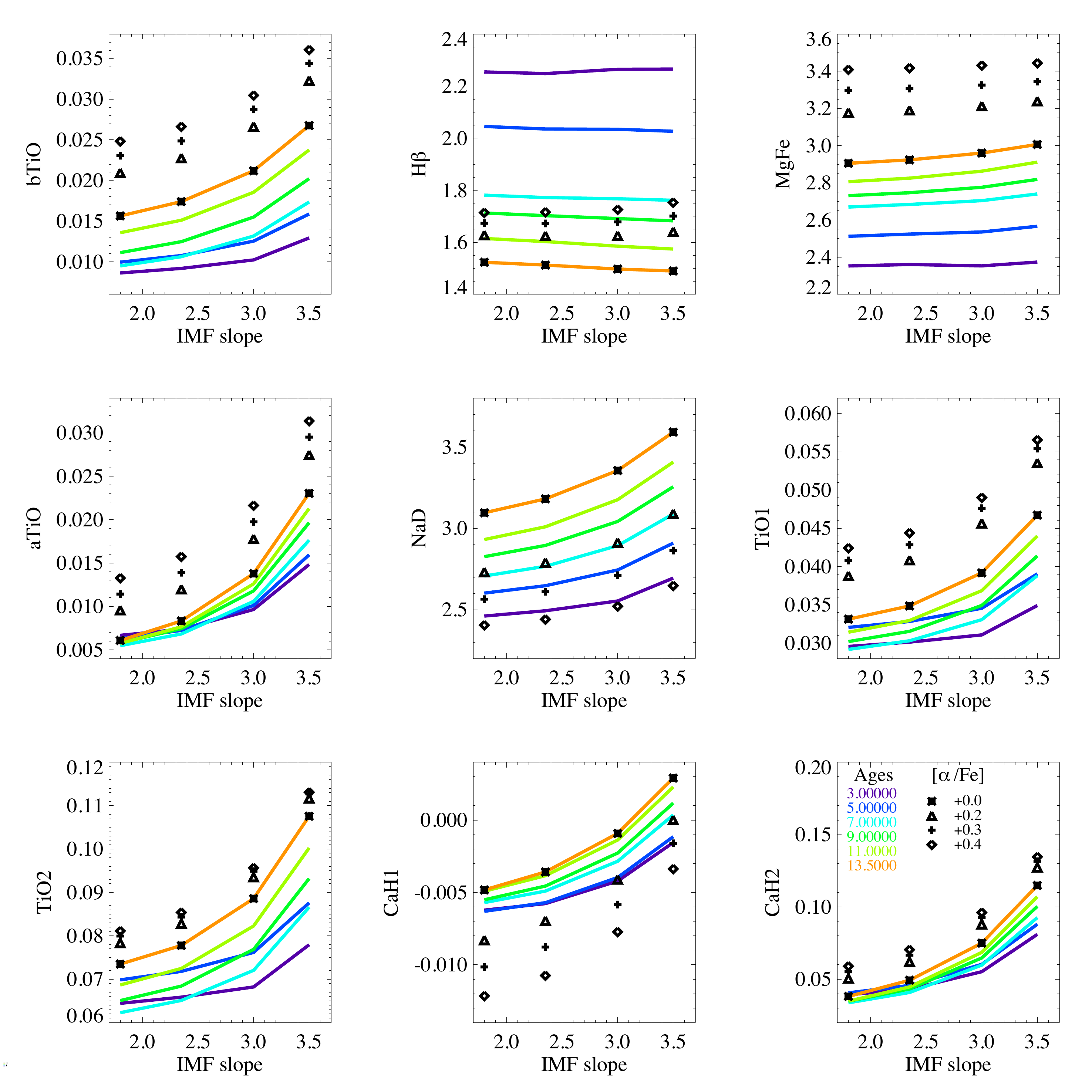}
\caption{Variation of index strengths with IMF slope for CvD12 models in the MILES wavelength 
range and with a resolution of $\sigma = 350\,\mathrm{km}\,\mathrm{s}^{-1}$. In each panel, different colors represent 
SSP models with different ages, while different symbols are SSP models with an age of $\sim13.5$ Gyr 
and with different [$\alpha$/Fe] ratios. The models predict a generally weak dependence of 
TiO features on age and an increase of all TiO strengths with increasing IMF. The CaH1 index 
strength also increases with steepening IMF and age but decreases with increasing [$\alpha$/Fe]. 
In all plots, a Salpeter IMF (\citealt{Salpeter1955}) has a slope of $x=2.35$.}
\label{fig:variationCONROY}
\end{figure*}

After the recent suggestion by vDC2010 that the low-mass end of the
IMF slope may be not universal and might depend on the stellar mass of the
system (\citealt{Treu2010, Spiniello2011, Spiniello2012, Cappellari2012}, CvD12), 
new SSP models with varying IMFs have been developed.  One of these is
CvD12, who presented SSP models with variable abundance patterns
and stellar IMFs suitable for studying the integrated-light spectra of
galaxies with ages $>3$ Gyr.  CvD12 synthesized stellar atmospheres
and spectra using the combination of three different isochrones to
describe separate phases of stellar evolution: the Dartmouth
isochrones (\citealt{Dotter2008}) for the main sequence and the red
giant branch (RGB); the Padova isochrones (\citealt{Marigo2008}) to
describe AGB evolution and the horizontal branch (HB); and the Lyon
isochrones (\citealt{Chabrier1997, Baraffe1998}) for the
lower-mass main sequence ($M_{\star}\leq 0.2\,M_{\odot}$).  All of the CvD12
models use {\sl solar} metallicity isochrones, even when synthesizing
with different abundance patterns or different $[\alpha/$Fe]. \\
\indent The models are based on empirical stellar libraries, modified 
using theoretical stellar atmosphere models and their emergent 
synthetic spectra, and therefore they are somewhat restricted in their SSP parameter coverage
(especially at high metallicity and for non-solar abundance ratios)
and poorly-calibrated against mass loss in advanced stages, such as
the asymptotic giant branch (AGB).  Stars in these phases provide a
non-negligible contribution to the red spectrum of galaxies (\citealt{Worthey1994}). 
This contribution is however hard to quantify because RGBs and AGBs 
are so short-lived that good statistics are hard to obtain from color-magnitude 
diagrams of globular and open clusters.  
Moreover, mass loss, which is a crucial
ingredient to calculate the precise evolutionary path of AGB, is very
hard to model and therefore is a main source of uncertainties in the
models (e.g.\ \citealt{Blocker1995}). \\
%CvD12 include a variety of nuisance
%parameters to take into account the contribution to the light from
%advanced evolutionary phases and they allow for the addition of
%arbitrary amounts of M giant light.  They also include a parameter to
%allow a ``frosting'' (\citealt{Trager2000}) of a young population with
%an age of 3 Gyr in addition to the age of the bulk population.\\
\indent The CvD12 models explore variations in age in the range 3--13.5 Gyr,
$\alpha-$enhancement of 0--0.4 dex, individual elemental abundance
variations and four different IMFs: a bottom-light Chabrier (2003) IMF, 
a Salpeter (1955) IMF with a slope of $x=2.35$ (where $x$ is the IMF slope, 
using $dN/dm \propto m^{-x}$),
and two bottom-heavy IMFs with slopes of $x=3.0$ and $x=3.5$.\\
\indent Figure~\ref{fig:variationCONROY} shows index strengths versus IMF slope
for different ages (different colors) and [$\alpha$/Fe] (symbols) for
a number of features from the CvD12 SSP models.  Most of the
  classical blue Lick indices (e.g.\ H$\beta$, [MgFe]: \citealt{Gonzalez1993, Worthey1994, Trager2000}) 
do not show a strong variation with the IMF slope, although they are important in
  constraining the age and metallicity of galaxies.  Index strengths of
the newly defined bTiO and aTiO and the redder TiO indices clearly
increase with steeper IMF slopes, but they also increase for higher
[$\alpha$/Fe]. 
On the other hand, the CaH1 index strength decreases with increasing
[$\alpha$/Fe]. This makes the combination of TiO lines and CaH1
absorption a very good IMF probe, as the effects of [$\alpha$/Fe] and varying IMF slope are orthogonal in a TiO-CaH plot.  

We note that other models, such as \citet{Thomas2011}, do not find a strong sensitivity of the TiO indices to  [$\alpha$/Fe].
The reason why previous models for TiO indices differ from CvD12 is that
the old response functions (including \citealt{Korn2005}, used in e.g., \citealt{Thomas2011}) 
do not include TiO molecules, so they will not be able to reproduce the TiO sensitivities. 

The dependence on age of index strengths is smaller for almost all the indicators if we restrict
ourself to ages $\geq 7$ Gyr.  A caveat of our analysis is that the
response functions of CvD12 do not currently include the CaH molecule. 
However observations of Ca indices indicate a nearly solar
[Ca/Fe] abundance in galaxies (\citealt{Vazdekis1997, Worthey1998, 
Trager1998, Cenarro2003, Smith2009}) and therefore 
this limitation in the response functions is unlikely to impact our results.

\section{Constraining the low-mass end of the IMF slope using optical spectra of ETGs}

We now compare the CvD12 models to early-type galaxies (ETGs), for which they are specifically designed.

\begin{figure*}  
\centering
\includegraphics[height=13cm]{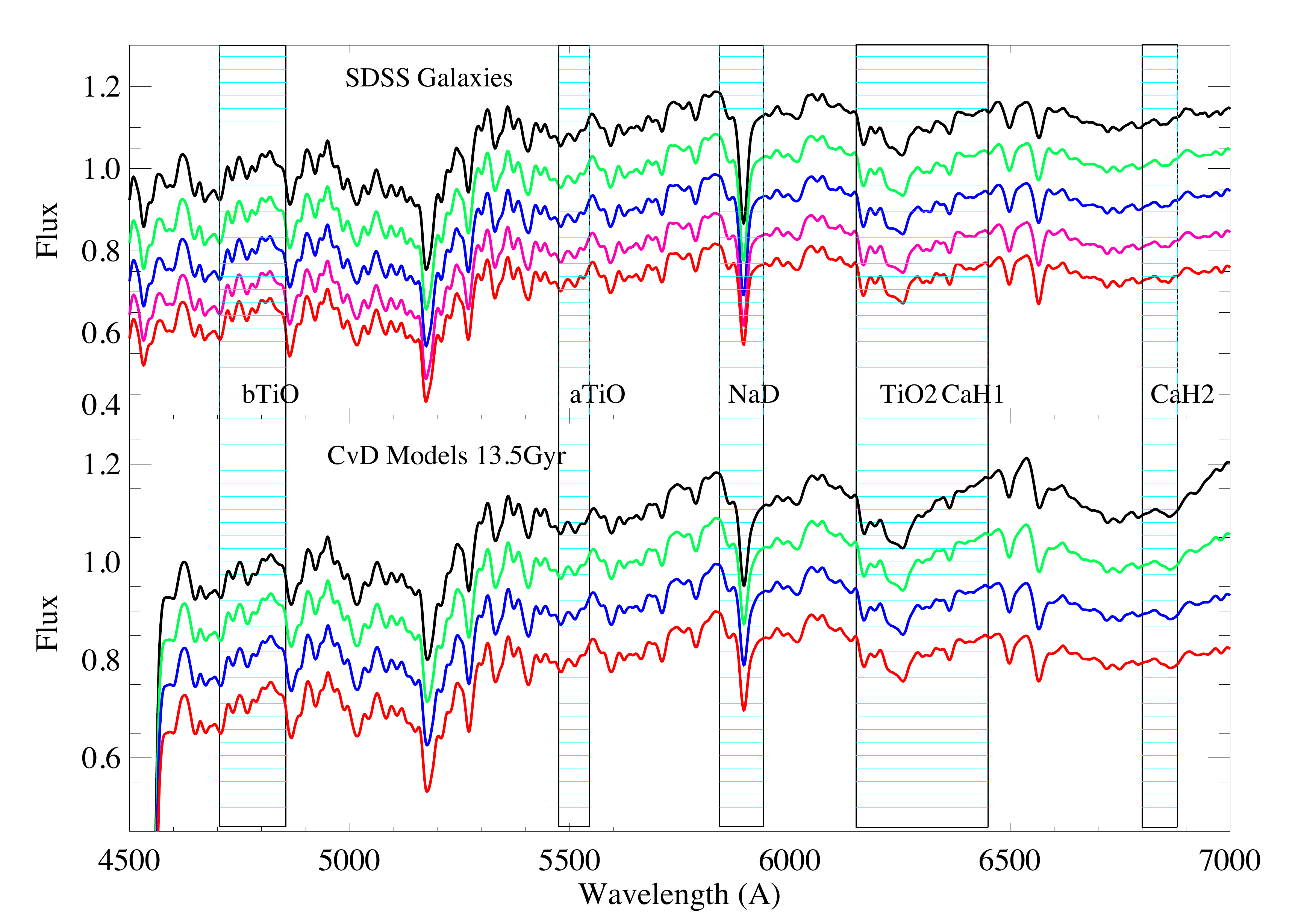}
\caption{ 
{\sl Upper panel:} Stacked spectra of SDSS galaxies with different observed
  velocity dispersions, increasing from bottom to top ($150$, $190$, $230$,$270$, $310\,\mathrm{km}\,\mathrm{s}^{-1}$ respectively) 
and offset for clarity. {\sl Lower panel:} CvD12 SSP models with same
  age (13.5 Gyr) and solar [$\alpha$/Fe] but different IMF slopes,
  from Chabrier (at the bottom) to the most bottom-heavy ($x=3.5$, at
  the top).  Boxes show IMF-sensitive features. All observed and model spectra are
  normalized with a polynomial fitting and convolved to a common
  resolution of $\sigma = 350\,\mathrm{km}\,\mathrm{s}^{-1}$.}
\label{fig:spectra}
\end{figure*}

We selected ETGs from the Sloan Digital Sky Survey DR8 (SDSS; \citealt{Aihara2011}) 
 in five velocity-dispersion bins spread over $150$--$310\,\mathrm{km}\,\mathrm{s}^{-1}$, each with a total 
width of $\sim 40\,\mathrm{km}\,\mathrm{s}^{-1}$.  To select ETGs and to minimize contamination of
  our sample by late-type galaxies, we select systems for which the
  galaxy's surface brightness profile has a likelihood of a de
  Vaucouleurs' model fit higher than the likelihood of an
  exponential model fit. This requirement reduces approximately to requiring that
   the surface brightness profile should be better fitted by the de Vaucouleurs' model than by an exponential. 
We also visually inspected spectra of all the selected galaxies removing that showing emission lines
    (specifically, all the cases in which $H_{\beta}$, O[ III] or O[ II] emission is visible). 
   Moreover, we select systems with very low
  star-formation rate (SFR $<0.3\,{M}_{\odot}\,\mathrm{yr^{-1}}$) using
  the MPA/JHU value-added galaxy catalog containing results from
  the galaxy spectral fitting code described in \citet{Tremonti2004} and star formation rates based on the technique
  discussed in \citet{Brinchmann2004}.
  Finally we set an upper limit on the redshift range ($z\leq 0.05$) to cover the entire wavelength
  range of interest.  In each bin we stack spectra together in order
  to increase the final signal-to-noise (S/N), which is of
  the order of $\sim 300/$\AA\, over the wavelength range
  4000--7000 \AA.
We use CvD12 models with varying IMF slopes, ages = [$7$--$13.5$] Gyr,
[$\alpha$/Fe] = [0, 0.2], and solar abundances.  We convolve the
galaxy and model spectra to an effective velocity dispersion of
$\sigma = 350\,\mathrm{km}\,\mathrm{s}^{-1}$ to correct for kinematic broadening before
measuring indices.  We also normalize the spectra using a second-order
polynomial fit.  In Figure~\ref{fig:spectra}, we show the stacked
SDSS spectra for each velocity dispersion bin as well as a 
set of models with solar [$\alpha$/Fe] and varying IMF slope. 
Boxes show the IMF-sensitive indices used in this work.

We measure line-strength indices in the range 4000--7400 \AA,
including the standard Lick indices H$\beta$, Mg$b$, Fe5270, Fe5335,
NaD and TiO2 (e.g., \citealt{Trager1998}), and the commonly-used [MgFe]
combination\footnote{$\mathrm{[MgFe]} = \sqrt{(\mathrm{Fe5270} +
    \mathrm{Fe5335})/2 \times \mathrm{Mg}b}$, \citet{Gonzalez1993}}, as
well as our newly defined IMF-sensitive features bTiO, aTiO and
CaH1.  Indices in both the galaxy and the model spectra are measured
with the same definitions and methods\footnote{We use the code SPINDEX
  from \citet{Trager2008}}.  We do not place our indices on the
zero-point system of the Lick indices and quote the new TiO
  and CaH indices as index strengths in units of magnitudes. Index-measurements for the stacked spectra of SDSS 
galaxies in the five velocity-dispersion bins are reported in Table~\ref{tab:ind_ssds_res}.

\begin{table*}
\caption{Equivalent widths of the features used in the following analysis for the stacked spectra of SDSS galaxies 
in the five velocity-dispersion bins with a total width of 
$40\mathrm{km}\,\mathrm{s}^{-1}$ ($\Delta \sigma = \pm 20\mathrm{km}\,\mathrm{s}^{-1}$). 
The bins have similar number of galaxies: 51, 54, 50, 49, 43 
from the lowest to the highest velocity-dispersion bin.   
The first five line-indices ($H_{\beta}$,Mg$b$, Fe5270,Fe5335, and NaD) are measured in angstrom, the others in magnitudes.} 
\centering 
\begin{tabular}{cccccc} 
\hline
\hline 
\\
{  \bf Index}  & { \bf $EW_{\sigma_{\star}=150\mathrm{km}\,\mathrm{s}^{-1}}$} & {\bf $EW_{\sigma_{\star}=190\mathrm{km}\,\mathrm{s}^{-1}}$ } & { \bf $EW_{\sigma_{\star}=230\mathrm{km}\,\mathrm{s}^{-1}}$ } & { \bf $EW_{\sigma_{\star}=270\mathrm{km}\,\mathrm{s}^{-1}}$ } & { \bf $EW_{\sigma_{\star}=310\mathrm{km}\,\mathrm{s}^{-1}}$}\\
\\
\hline
\\
$H_{\beta}$ &		$1.666  \pm  0.001  $   	&		$1.523 \pm 0.001$		&	$1.449 \pm 0.004$			&	$1.496 \pm 0.003$	&	$1.491\pm 0.008$\\
Mg$b$			&		$3.531  \pm  0.001  $	 	&		$3.878 \pm 0.001 $		&	$4.088 \pm 0.004 $			&	$ 4.155 \pm 0.003$	&	$ 4.282\pm 0.008$\\
Fe5270		    &		$2.364  \pm  0.001  $		&		$2.389 \pm 0.001 $		&	$2.400 \pm 0.005$			&	$2.466 \pm 0.004$	&	$2.499 \pm 0.010$\\
Fe5335		    &		$1.768  \pm  0.001  $		&		$1.784 \pm 0.002$ 			&	$1.889 \pm 0.006$			&	$ 1.803 \pm 0.005$	&	$1.895\pm 0.012$\\
NaD			    &		$3.308 \pm   0.001  $		&		$3.9238 \pm 0.0001$			&	$4.413 \pm 0.003 $		&	$4.628\pm 0.003$	&	$ 4.962\pm 0.007$ \\
\\
bTiO			    &		$\hphantom{-} 0.01589  \pm 0.00004	$  &		$\hphantom{-} 0.01936 \pm 0.00004$		&	$\hphantom{-} 0.0205 \pm 0.0002$		&	$\hphantom{-} 0.0207 \pm 0.0001$	&	$\hphantom{-} 0.0212\pm 0.0003 $\\
aTiO			    &		$\hphantom{-} 0.00839  \pm 0.00002$	&		$\hphantom{-} 0.01153 \pm 0.00003$		&	$\hphantom{-} 0.0126 \pm 0.0001$		&	$\hphantom{-}  0.0114\pm 0.0001$	&	$\hphantom{-} 0.0387\pm 0.0002$ \\
TiO1			    &		$\hphantom{-} 0.03141  \pm 0.00002$	&		$\hphantom{-} 0.03664 \pm 0.00002$		&	$\hphantom{-} 0.0390 \pm 0.0001$		&	$\hphantom{-} 0.39913\pm 0.0001$	&	$\hphantom{-} 0.0387\pm 0.0002$\\
TiO2			    &		$\hphantom{-} 0.07508  \pm 0.00002$	&	 	$\hphantom{-} 0.0815 \pm 0.00002$	&	$\hphantom{-} 0.08364 \pm 0.0001$		&	$ \hphantom{-} 0.0880\pm 0.0001$	&	$\hphantom{-} 0.0900\pm 0.0002$\\
CaH1		    &		$-0.00532 \pm 0.00002$ &		$-0.00467 \pm 0.00003$&	$ -0.0042 \pm 0.0001$	&	$ -0.0033\pm 0.0001$	&	$  -0.0321\pm 0.0002$\\
CaH2		    &		$\hphantom{-} 0.03471   \pm 0.00002$  &		$\hphantom{-} 0.03503 \pm 0.00002  $	&	$\hphantom{-} 0.0368 \pm 0.0001$		&	$\hphantom{-} 0.0349\pm 0.0001$	&	$\hphantom{-} 0.0359\pm 0.0002$ \\
\\
\hline 
\\
\hline
\hline 
\end{tabular} 
\label{tab:ind_ssds_res}
\end{table*}

\begin{figure*}
\centering
\includegraphics[height=16cm]{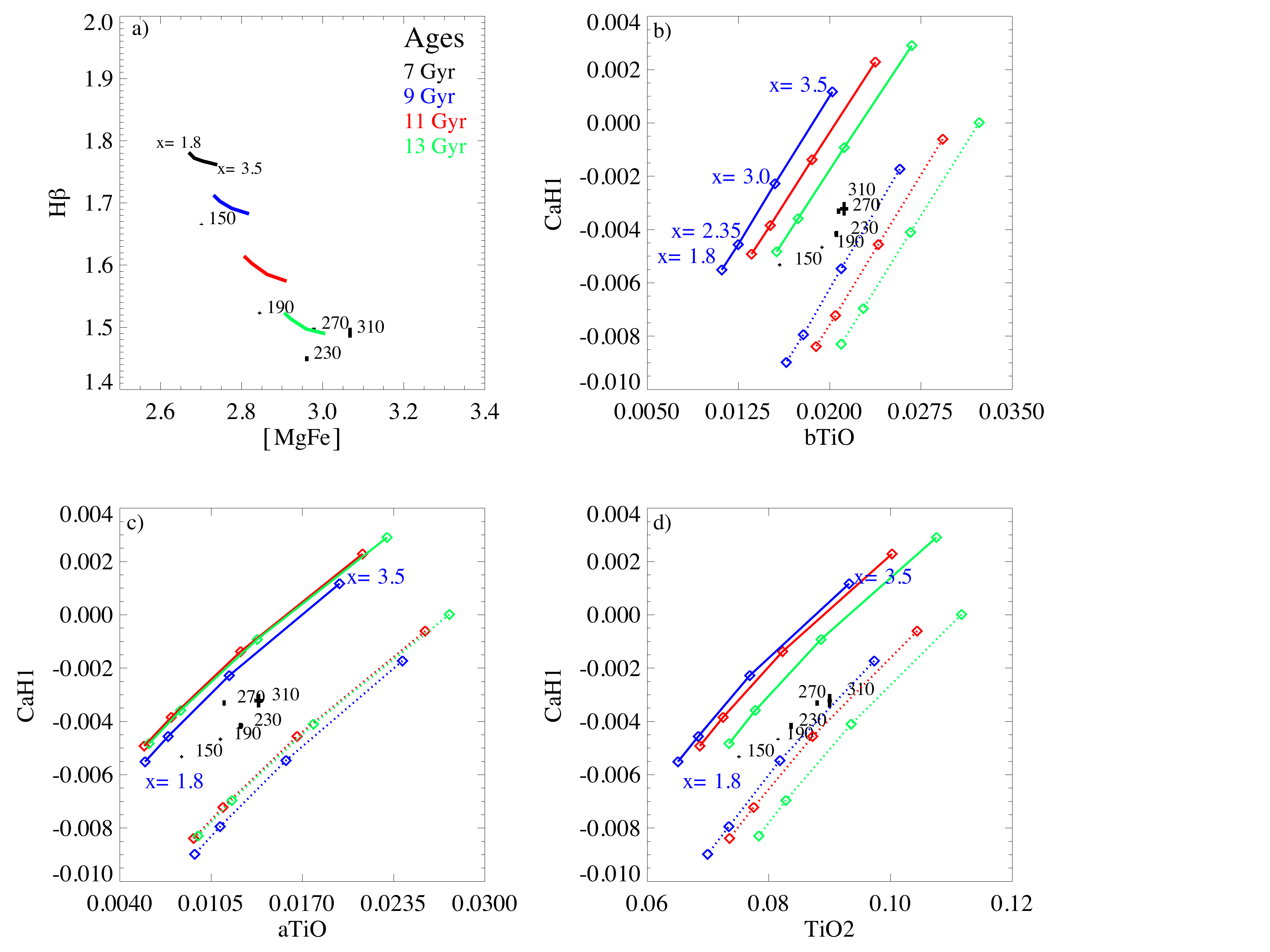}
\caption{Index-index plots of the main absorption features in the
  optical. Solid lines are CvD12 models with different ages (colors)
  and different IMF slopes (points on each line are models with IMF
  slopes of 1.8, 2.35, 3.0, 3.5). Dotted lines are the same models
  with $[\alpha/\mathrm{Fe}]=+0.2\,\mathrm{dex}$. All models have
  solar metallicity (i.e., [Fe/H]=0.0).  Points with error bars are stacked spectra of
  SDSS galaxies, in five velocity-dispersion ($\sigma_{\star}$) bins.  The
  value shown is the average $\sigma_{\star}$ in $km\,s^{-1}$ for each bin.
  Galaxies and models are convolved to a final common resolution of
  $350\,\mathrm{km}\,\mathrm{s}^{-1}$ before measuring index strengths.  \textit{Panel (a):}
  The H$\beta$--[MgFe] diagram. In this plot IMF dependence is
  minimal, and the ages and metallicities of the galaxies can be
  inferred. All the galaxies show populations older than 9 Gyr.
  \textit{Panels (b, c, d):} IMF-sensitive index-index plots; CaH1 as
  a function of bTiO, aTiO and TiO2 respectively.  
  The TiO lines are strong in cool giants and dwarfs and become strong with increasing  [$\alpha$/Fe]. 
CaH1 comes only from M dwarfs and becomes weaker for super-solar [$\alpha$/Fe]. 
The combination of these indices is therefore particularly effective at constraining the 
low-mass end of the IMF slope.  
  All the indicators
  agree and confirm a steepening of the IMF slope with galaxy mass. In
  these panels, galaxies prefer models with slightly super-solar
  [$\alpha$/Fe].  }
\label{fig:sdss}
\end{figure*}

To investigate the IMF variation with galaxy mass, we produce
index-index plots for the selected optical indicators.  Panel (a) of
Figure~\ref{fig:sdss} shows the H$\beta$--[MgFe] diagram, used to
constrain age and metallicity of galaxies.  The different colors
represent CvD12 SSP models with different ages and solar
[$\alpha$/Fe].  We observe an increase in age with increasing velocity
dispersion, consistent with previous stellar population studies of
early-type galaxies that suggest that more massive galaxies
predominantly have older, more evolved, stellar populations  (e.g.\ \citealt{Renzini2006}).  
The reader might notice that contrary to many other SSP models (\citealt{Worthey1994, Bruzual1993, Bruzual2003, Leitherer1999, Vazdekis1996, Vazdekis1997,Maraston2005, Thomas2005}), in the CvD12 models the [MgFe] index depends on [$\alpha $/Fe].
This is because in the CvD12 models the abundance variations of single elements are implemented at fixed [Fe/H]\footnote{
\citet{Schiavon2007} has already seen the same effect in his models which have also been cast in terms of [Fe/H], and not total metallicity [Z/H]. This reflects the choice to deal explicitly with quantities that can be inferred from measurements taken in the integrated spectra
of galaxies. Total metallicity is not one of them, given our current inability to use integrated 
spectra of stellar populations to constrain the most abundant of all metals, oxygen. As clearly discussed in \citet{Graves2008}, 
an advantage with casting models in terms of [Fe/H] is that each elemental abundance
can be treated separately, so that the effect of its variation can be studied in isolation from
every other elemental abundance at the cost, however, of varying the total metallicity. In the case
of models cast in terms of [Z/H], it is impossible to vary the abundance of a single element,
because enhancing one element means decreasing the abundances of all other elements to keep the 
total metallicity constant.}.
All selected galaxies are consistent with populations of 9
Gyr or older, and therefore in the other panels we only plot SSP
models with ages of 9, 11, 13.5 Gyr.  We plot models with solar
[$\alpha$/Fe] (solid lines) and supersolar [$\alpha$/Fe]
($=0.2\,\mathrm{dex}$, dotted lines); we note that the predicted
variation of IMF slopes is orthogonal to the [$\alpha$/Fe] enrichment
for the selected IMF-sensitive indicators (\citealt{Spiniello2012}). 
Panels (b), (c) and (d) of Figure~\ref{fig:sdss} show some of our
selected IMF indicators in the optical. 
In each panel we plot one of the TiO lines against CaH1, which is the only feature that is strong only in cool dwarfs. 
The combination of TiO lines (whose strengths increase with steepening of the IMF 
{\sl and} with increasing [$\alpha$/Fe]) and CaH1 absorption (whose index strength 
instead decreases with increasing [$\alpha$/Fe]) allows us to break the IMF 
slope--[$\alpha$/Fe] degeneracy. 
Several TiO lines are necessary to further break the IMF--$T_{\rm eff, RGB}$, as described below. 

Remarkably, all indicators clearly show a steeper IMF
slope with increasing galaxy mass: they imply a Chabrier IMF for the
least-massive galaxies ($\langle \sigma_{*}\rangle \sim 150\,\mathrm{km}\,\mathrm{s}^{-1}$),
Salpeter for the intermediate-mass galaxies ($\langle \sigma_{*}\rangle \sim 230\,\mathrm{km}\,\mathrm{s}^{-1}$) 
and possibly a bottom-heavy IMF with $x \la 3$ for the most massive SDSS ETGs
($\langle \sigma_{*}\rangle \sim 310\,\mathrm{km}\,\mathrm{s}^{-1}$). We quantify this further in section 4.3.

\subsection{Variation in the $\Delta {T_{\rm eff}}$ of the red-giant branch versus IMF variations}

\begin{figure*}
\centering
\includegraphics[height=12.5cm]{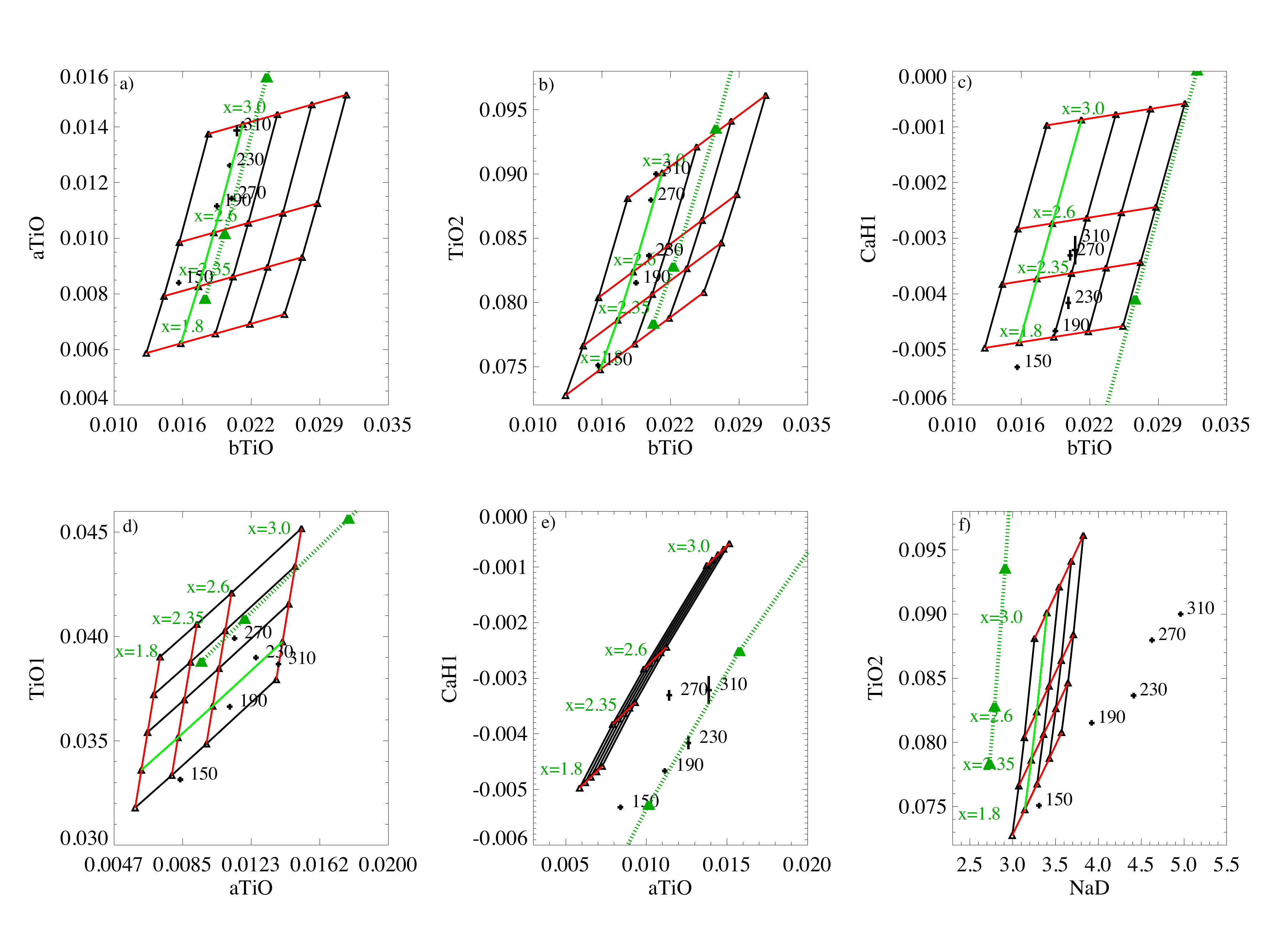}
\caption{ The effect of $\Delta T_\mathrm{eff,RGB}$ on CvD12 models with ages of 13.5 Gyr. 
In each panel, the green solid line shows the standard-temperature model with solar [$\alpha$/Fe] while the
  black lines are models in which $\Delta {T_{eff, RGB}}$ has been increased or
  decreased by $50\,\mathrm{K}$. 
  The green dotted line shows the standard-temperature model with [$\alpha$/Fe]$=+0.1$. 
   We plot models with $\Delta {T_{eff, RGB}} =
  50\,\mathrm{K}$, $-50\,\mathrm{K}$, $-100\,\mathrm{K}$ and
  $-150\,\mathrm{K}$. Red lines represent therefore $\Delta {\rm T_{\rm eff, RGB}}$
  variations, while different symbols on the same black line indicate
  different IMF slopes for a model with a given temperature. Points
  with error bars are stacked spectra of SDSS galaxies, in five
  velocity-dispersion ($\sigma_{\star}$) bins whose average is shown in
  $km\,s^{-1}$.  {\sl Panels a--d):} The combination of these indicators
  allow the degeneracy between RGB temperature and IMF to be broken. A
  mild variation of $\Delta {T_{\rm eff, RGB}}$ with galaxy mass can be detected,
  but a clear trend of increasing IMF with galaxy mass is present in
  all three panels. {\sl Panel e):} CaH1--aTiO does not allow
  simultaneous constraints on the IMF slope and $\Delta {T_{\rm eff, RGB}}$. 
  From this panel it is clear that models with slightly super-solar [$\alpha$/Fe] better fit the data. 
  The same result is also visible in Figure~\ref{fig:sdss} (c) and (d) and is consistent 
  with the best-fit models of Table~\ref{tab:imf_sdss}.  
  {\sl Panel f):} NaD strengths in the stacked galaxies show a strong
  variation with galaxy mass. This variation cannot be entirely
  explained via non-universality of the IMF  (\citealt{Spiniello2012}). 
  The data suggest that NaD is highly dependent on
  the temperature of the RGB. }
\label{fig:IMF_Teff}
\end{figure*}

Here we investigate further the IMF--$\Delta {T_{\rm eff, RGB}}$
degeneracy, showing that the data are consistent with a non-universal
IMF even when we take into account the possible uncertainty on the
exact temperature of the red giant branch in ETGs.  We use 13.5 Gyr old CvD12 models in which we change 
the temperature of the RGB in the isochrones.  In each index-index plot of
Figure~\ref{fig:IMF_Teff} we show the standard-temperature model with solar and super-solar [$\alpha$/Fe], 
a solar-abundance model with $\Delta {T_{\rm eff , RGB}}$ increased by $50\,\mathrm{K}$,
and three solar-abundance models with $\Delta {T_{\rm eff, RGB}}$ decreased by
respectively $50\,\mathrm{K}$, $100\,\mathrm{K}$, and
$150\,\mathrm{K}$. On each line, different symbols indicate different
assumed IMFs.  Figure~\ref{fig:IMF_Teff} clearly shows that 
the combination of our preferred optical IMF-dependent indicators
allows us to break also the IMF-$\Delta {T_{\rm eff, RGB}}$ degeneracy. 
It is also clear from the figure that ETGs are slightly $\alpha$-enhanced, and that
the degeneracy between IMF, $\Delta {T_{\rm eff, RGB}}$ and [$\alpha$/Fe] 
can be completely broken by using several TiO lines in combination with the CaH1 feature.

In principle the NaD index at 5900~\AA\, could be used to constrain the IMF 
(\citealt{Conroy2012b, Ferreras2013}), but, as shown in panel (f) of Figure~\ref{fig:IMF_Teff}, this index is more sensitive 
to variation of the effective temperature of the RGB stars than to variation of the IMF slope in these models. 
However, this feature is above all sensitive to [Na/Fe] abundance, as we will show in the next section.

\subsection{The Na features}
\label{section:nad}
NaD and NaI indices are both very strong absorption features in the optical-NIR 
spectrum and have been the subject of numerous
studies of the IMF slope as well as of the interstellar medium (ISM). 
The first claim of a non-universal low-mass end of the IMF was made by \citet{Spinrad1971} 
using the NaI feature ($\lambda \sim8190$ \AA).
They compared the observed line strength of the NaI doublet in the centers of M31, M32 and
M81 with population synthesis models, finding a substantial fractional contribution by dwarf stars to the
integrated light of these galaxies.
More recently, \citet{vandokkum2010} demonstrated indeed that the NaI doublet depends strongly on surface 
gravity at fixed effective temperature, betraying the presence of faint M dwarfs in integrated light spectra. 
Figure~\ref{fig:miles2} shows that the bluer NaD feature ($\lambda \sim5900$ \AA) is also 
very strong in dwarf stars at relatively low temperatures ($T_{\rm eff} \le 4200$K). 
Hence, its strength as a function of the redder sodium feature should provide a powerful means 
for separating the IMF from the sodium abundance, assuming that the two features react in 
the same way to changes in the [Na/Fe] abundance.
However, it has been argued that the NaD index could be highly contaminated by the ISM, even though 
the dominant process is under great debate (cool gas in the disk, galactic wind in 
actively star-forming galaxies or AGN activity are the most common candidates, 
e.g.\ \citealt{Heckman1990, Lehnert2011}).

Here we show that the NaD strength is not {\sl solely} driven by IMF variation: 
we find that this index is more sensitive to [Na/Fe] abundance variations 
than to IMF slope variations (Spiniello et al. 2013b, in prep.), and it varies strongly with $T_{\mathrm{eff,RGB}}$. 
In \citet{Spiniello2012} we show that CvD12 models with solar [Na/Fe] 
abundance do not match the NaD strengths for high-mass ETGs. This is also confirmed by 
\citet{Conroy2012b} who find that massive ETGs require models with $\mathrm{[Na/Fe]}=0.3$--$0.4$. 
When {\sl only} sodium indicators are used these systems require also extremely steep IMFs, $x\ge 3.5$.
For the gravitational lens analysed in \citet{Spiniello2012} paper (with $\sigma_{\star} \sim 300\,\mathrm{km}\,\mathrm{s}^{-1}$), 
such steep slopes are ruled out at the 99.9 per cent confidence level (C.L.) from lensing constraints. 
We therefore conclude that Na lines are highly sensitive to both IMF variation and [Na/Fe] variation, 
and are therefore complicated to interpret when considered in isolation.  

%If index strengths are to be used to analyse spectra of massive ETGs, 
Furthermore, in a recent paper, \citet{Smith2013} found a giant elliptical ($\sigma_{\star} \sim 330\,\mathrm{km}\,\mathrm{s}^{-1}$) 
lens galaxy with a ``lightweight" IMF. From their lensing analysis, an IMF heavier than Salpeter 
is disfavoured at the $> 99.8$ per cent level. 
However, when looking at the dwarf-sensitive NaI feature in the spectrum of that galaxy, a SSP with 
age of $13.5$ Gyr, super-solar [$\alpha$/Fe] {\sl and} bottom-heavy IMF ($x=3.0$) is required 
to match the observed galaxy spectrum. Also in this case, the stellar mass inferred from this SSP model violates 
the limits on the total mass of that particular system, set by gravitational lensing.  
Even when considering models with an enhancement of sodium, \citet{Smith2013} cannot completely reconcile the 
SSP Na-based result and the lensing result on the IMF slope {\sl for this} particular giant elliptical. 

Additional evidence showing that the NaD strengths in ETGs are not  
exclusively IMF-driven is provided by Jeong et al. (2013). 
These authors identified NaD excess objects (NEOs) from the SDSS DR7 (\citealt{Abazajian2009}) and found that NaD 
features are too strong in many massive ETGs even when using SSP models with a bottom-heavy IMFs (and solar abundance).
They studied a plausible range of stellar parameters (from CvD12 SSP models) that could reproduce 
the observed values of NaD, Mg$b$ and Fe5270, finding that the majority of the early-type 
NEOs are $\alpha$-enhanced, metal-rich and especially Na-enhanced ([Na/Fe] $\sim 0.3$).

To further complicate this situation, we note that large differences exist when comparing the behaviour of 
Na features predicted from different SSP models, as pointed out in \citet{LaBarbera2013}.
The authors compared the [$\alpha$/Fe] sensitivity of the NaD for  
different SSP models finding a strong discrepancy. 
They showed that the CvD12, \citet{Coelho2007} and \citet{Cervantes2007} 
models predict that NaD weakens as [$\alpha$/Fe] 
increases, while the \citet{Thomas2011} models predict a mild increase with
enhancement.

Given this unresolved situation about the level of sodium abundance in ETGs, we therefore think that 
it is currently prudent not to draw any strong conclusion about the low-mass end of the IMF 
slope based {\sl solely} on Na features, and to concentrate on indices such as those from  
TiO and CaH, which appear to be less subject to abundance-ratio 
variation effects and give mutually consistent predictions for the trend of IMF slope with galaxy mass. 

In the following section we give a quantitative expression for the relation between 
IMF slope and velocity dispersion, showing that the relation inferred from sodium lines is 
systematically different than the relation inferred from any other combination 
of the above-mentioned IMF-sensitive spectral indices.

\subsection{IMF slope versus velocity dispersion}

Our results confirm that it is possible to constrain the low-mass end
of the IMF slope from optical spectra of ETGs and provide increasing
support to the emerging notion of the non-universality of the IMF. We observe that the 
steepness of the low-mass end of the IMF 
 increases with galaxy velocity dispersion, consistent with previously-published
works (\citealt{vandokkum2010, Spiniello2011, Spiniello2012, Conroy2012b, Cappellari2012, Ferreras2013, Tortora2013,LaBarbera2013}). 
To constrain the IMF slope of each galaxy and
concurrently break the degeneracies between age, metalicity, abundance ratio and
temperature, we use the following indices: H$\beta$, Mg$b$, Fe5270,
Fe5335, bTiO, aTiO, TiO1, TiO2, CaH1, and CaH2.   
The first four indicators are almost IMF-independent and they are mainly used to constrain age and [$\alpha$/Fe], while the 
combination of CaH1 with TiO features is used to count dwarfs stars and 
to break the degeneracy with effective temperature (see Figs.~\ref{fig:sdss} and \ref{fig:IMF_Teff}).
We also test the effect of including or not the NaD index.

We now give a quantitative expression for the variation of the IMF
slope with velocity dispersion.  We compare each stacked SDSS spectrum
with grids of interpolated SSPs spanning a range of ages (${\log({\rm age})} = [0.8
  - 1.15]$ Gyr, with a step of $0.01$ Gyr), [$\alpha$/Fe] (between
$-0.2$ and $+0.4$ dex, with a step of 0.05 dex), and changes in the
effective temperature of the RGB ($\Delta {T_{\rm eff, RGB}}$ between
$-200\,\mathrm{K}$ and $50\,\mathrm{K}$, with a step of
$50\,\mathrm{K}$) for different values of the IMF slope ($x= 1.8 -
3.5$, with a step of 0.1 in the slope).  
Using a bspline interpolation, we build a grid of $8 \times
13 \times 18 \times 9$ models in ${\log({\rm age})}$, [$\alpha$/Fe],
IMF slope and $\Delta {T_{\rm eff, RGB}}$, and for each galaxy
spectrum we compute the $\chi^2$ values for different  sets of optical indicators
as
\begin{equation}
\chi^2_{n} = \sum_{ind = 1}^{k} \chi^2_{ind, n}= \sum_{ind = 1}^{k}
\, \dfrac{(EW^{obs}_{ind} \,-\,EW^{mod}_{ind, n})^{2}}{\sigma^{2}_{EW^{obs}_{ind}}},
\end{equation}
where $n$ is the SSP model of interest and $k$ is the number of indices.
We obtain a probability density function (PDF) for each model via the
likelihood function $L \propto \exp( -\chi^{2}/2)$.  We then
marginalize over age, $\Delta {T_{eff}}$ and [$\alpha$/Fe] to obtain
a best-fitting slope of the IMF and its uncertainty ($1\sigma$ error
on the cumulative probability distribution) for each velocity
dispersion bin, assuming flat priors on all parameters.  

\begin{figure}
\includegraphics[height=7.2cm]{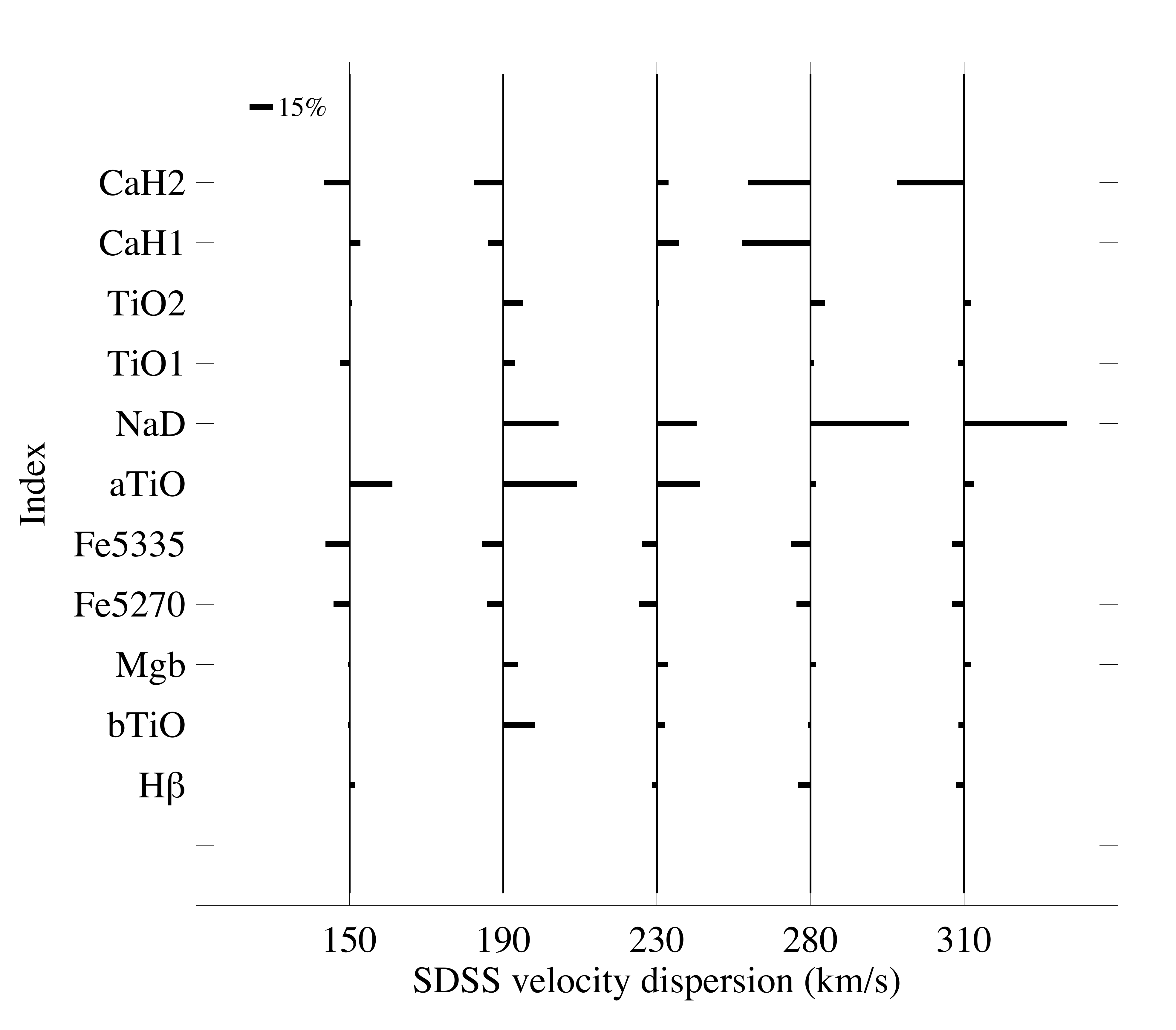}
\centering
\caption{Deviation of the EWs measured in the stacked SDSS galaxy spectra ($EW_{\rm SDSS}$) from the 
values measured in the corresponding SSP best-fit model ($EW_{\rm BF}$). 
For each velocity dispersion bin, shown on the $x$-axis, the vertical lines indicate the model values of the fitted indices, 
indicated on the $y$-axis. Horizontal lines departing from these indicate the percentile variations of the given index. 
As reference, we show the length of a line corresponding to a deviation of 15 per cent on the upper-left corner. 
When the value of the $EW_{\rm SDSS}$ is bigger than the value of the $EW_{\rm BF}$, the horizontal line goes to the right, as for 
NaD. The $EW_{\rm SDSS}$ of the NaD feature in the more massive galaxies are systematically 
underestimated by the SSP models, because we are restricting the analysis to solar [Na/Fe] (see Sec.~\ref{section:nad}).}
\label{fig:bestfit_SSP}
\end{figure}

We present our SSP analysis results in Table~\ref{tab:imf_sdss}. 
We plot the percentile deviation of the EWs measured in the stacked SDSS galaxy spectra ($EW_{\rm SDSS}$) from the 
values measured in the correspondent SSP best-fit model ($EW_{\rm BF}$) and Figure~\ref{fig:bestfit_SSP}. 

\begin{table*}
\caption{
Best-fit stellar population models for SDSS galaxies in different velocity dispersion bins. 
$\Delta T_{\rm eff}$ shifts are given with respect to the effective temperature of the RGB of solar-scaled models.} 
\centering 
\begin{tabular}{ccccl} 
\hline
\hline 
{  \bf SDSS $\sigma_{\star}$}  & {\bf  Best-fit age } & { \bf Best-fit [$\alpha$/Fe] } & { \bf Best-fit IMF slope } & { \bf Best-fit $\Delta T_\mathrm{eff}$}\\
{\bf ($km\,s^{-1}$)}  & {\bf (Gyr) } &   & { \bf (Salpeter: $x=2.35$)} & { \bf (K)}\\
[0.5ex]	% inserts table %heading \hline 1&50&837&970 \\ 2&47&877&230 \\ 3&31&25 &415 \\ 4 & 35 & 144 & 2356 \\ 5 & 45 & 300 & 556 \\ [1ex] 
\hline 
$150 \pm 20$ & $10 \pm 2 $ & $-0.11 \pm 0.04$ & $1.85 \pm 0.2$ &  $- 50  \pm 30$\\ 
$190 \pm 20$ & $11 \pm 2 $ & $+0.05 \pm 0.02$ & $2.08 \pm 0.2$ &  $-98 \pm 36$\\ 
$230 \pm 20$ & $13 \pm 2 $ & $+0.11 \pm 0.02$ & $2.33 \pm 0.4$ &   $-115 \pm 29$\\
$270 \pm 20$ & $12 \pm 2 $ & $+0.16 \pm 0.04$ & $2.40 \pm 0.3$ &   $-80 \pm 30$\\
$310 \pm 20$ & $13 \pm 3 $ & $+0.14 \pm 0.02$ & $2.62 \pm 0.4$ &   $-131 \pm 40$\\
\\
\hline
\hline 
\end{tabular} 
\label{tab:imf_sdss}
\end{table*}

In Figure~\ref{fig:IMF_sigma} we show the best-fit
IMF slope as a function of central stellar velocity
dispersion. We determine the parameters of a linear
  regression model, $x = a \, \times \, \log\,\sigma_{200} + b$, for
  the stacked SDSS spectra to be
\begin{equation}
x = (2.3\pm 0.1) \, \log\,\sigma_{200} + (2.13 \pm 0.15) \, ,
\end{equation}
when NaD is not included and 
\begin{equation}
x = (2.6\pm 0.2) \, \log\,\sigma_{200} + (2.3 \pm 0.2) \, ,
\end{equation}
when using NaD.
In the equations $x$ is the IMF slope and $\sigma_{200}$ is the central stellar
velocity dispersion measured in units of $200\,\mathrm{km}\,\mathrm{s}^{-1}$. 

To test the robustness of the IMF--$\sigma$ relation, we repeat the
above statistical procedure several times eliminating each
time one or more indices and calculating a new likelihood function for
each model and the inferred stellar population parameters for each
stacked spectrum.  
We always keep $H\beta$, our only age-dependent and IMF-independent feature. 
Our inferred IMF slopes are consistent within $1\sigma$ if we use CaH1, 
$H\beta$, Mg$b$, at least one of the iron lines, one of the blue TiO indices (bTiO or aTiO) 
and one of the redder TiO features (TiO1 or TiO2). 
%We always include CaH1 because this is the only feature that comes almost exclusively from dwarfs, and we also keep
%H$\beta$, which is the only age-dependent feature.  
If we use the [MgFe] as a single indicator we still recover the IMF
variation with velocity dispersion, but we do not recover the
well-established [$\alpha$/Fe]--$\sigma_{\star}$ relation (\citealt{Trager2000, Arrigoni2010}). 
%If we remove Mg$b$ or Fe5270 and Fe5335, we are not able to constrain the [$\alpha$/Fe] abundance. 
If we do not include CaH1 we are not able to break the degeneracy between [$\alpha$/Fe] and IMF slope. 
In this case, all the IMF-sensitive features increase both with steepening of the IMF slope and increasing of [$\alpha$/Fe].
The aTiO index does not significantly affect the IMF slope if we include
or exclude it, confirming that the [OI] sky line does
not have a significant effect on the inference of the IMF slope from this absorption feature. 
However, if we exclude aTiO from the analysis we find RGB effective temperatures
systematically higher than those inferred by including aTiO. 

\begin{figure}
\includegraphics[height=6.5cm]{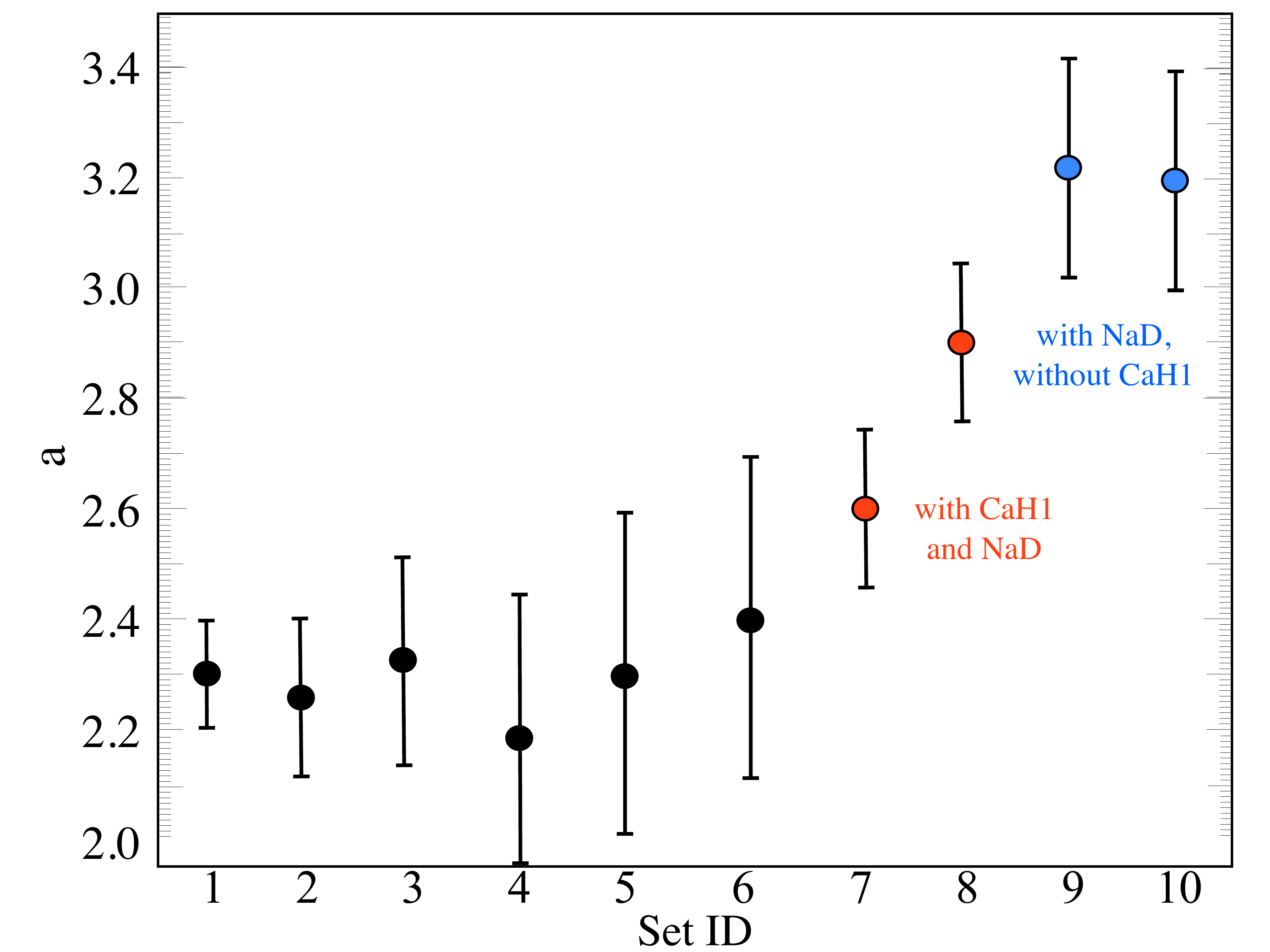}
\caption{The slope of the linear relation $x = a\, \log\,\sigma_{200} \,+ \,b$ inferred by 
using the different sets of indices as given in Table~\ref{tab:imf_sdss_indicator}. 
The results are robust and stable against systematics arising from single indicators. 
When using NaD we find a steep relation. }
\label{fig:leastsquare_a}
\end{figure}

\begin{figure*}
\includegraphics[height=7cm]{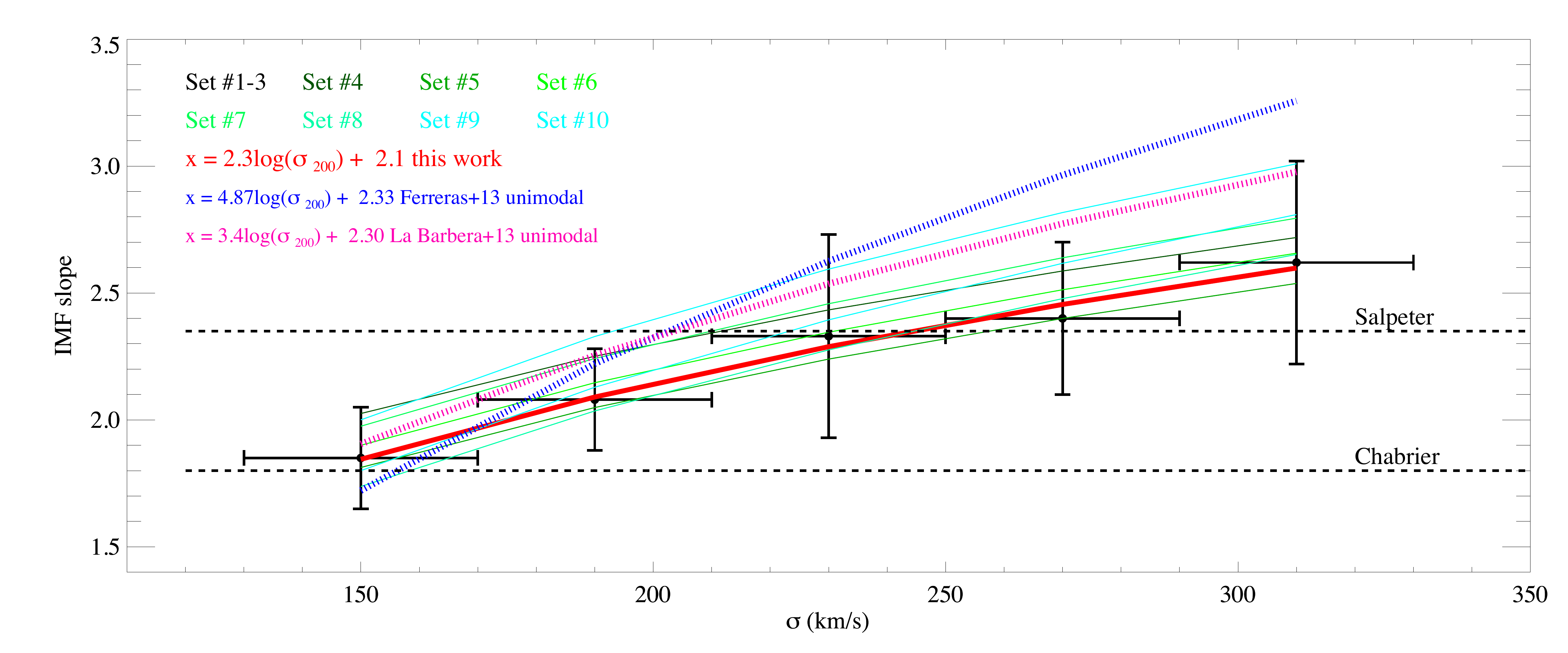}
\caption{Variation of the IMF slope as a function of stellar velocity
  dispersion. Points are SDSS ETGs stacked by velocity dispersion.
  The red solid line represents the the IMF-$\sigma_{200}$ relation obtained by using our preferred set of ten indicators,  
  as described in Table~\ref{tab:imf_sdss_indicator}. The colored lines show the same relation for each set of indicators. 
  The blue dotted line is the linear fit obtained by F13, 
 and the magenta dashed line is the linear fit obtained by LB13 for a unimodal IMF. 
 A very good agreement is found with LB13, when using a similar set of indicators (i.e. including NaD and excluding CaH1). 
  The Chabrier and Salpeter cases are shown as horizontal dashed lines.}
\label{fig:IMF_sigma}
\end{figure*}

The situation changes when NaD is included: 
the slope of the linear relation becomes significantly steeper when we include this index. 
In Figure~\ref{fig:leastsquare_a} we show this effect by plotting the resulting value of the slope 
of the linear relation $x = a\, \log\,\sigma_{200} \,+ \,b$ inferred from different sets 
of indicators noted in Table~\ref{tab:imf_sdss_indicator}. 
We note that we are restricting this particular analysis to solar Na abundances. 
A trend of increasing [Na/Fe] abundance with galaxy mass can {\sl partially} mimic the 
IMF steepening. We address this in a forthcoming paper.
It is important to note that we are assuming ${\rm [Na/Mg]}=0.0$, which may not be a good assumption 
for these very massive ETGs. A full spectral fitting approach as in \citet{Conroy2012b} and \citet{Conroy2013b} 
might be necessary to investigate in detail the effect of a single non-solar element abundance. 

We note a relation between $\Delta {T_{eff}}$ and $\sigma$ in 
Table~\ref{tab:imf_sdss}: stacks of the spectra of low-$\sigma$ galaxies show smaller temperature shifts than 
stacked spectra of high-$\sigma$ galaxies. 
We believe that the basis for this trend is to be found in a trend between $\Delta {T_{eff}}$ 
and metallicity or more specifically the overall abundance of $\alpha$-elements
 (i.e., \citealt{Salaris1993, Pietrinferni2006, Dotter2007}).

To test whether the trend $\Delta {T_{eff}}$ -- $\sigma$ trend has an impact on the IMF--$\sigma$ 
relation, we repeat the same analysis by fixing $\Delta {\rm
T_{\rm eff, RGB}}=0K$ and $=-100K$ for the set ID 4 and 9, which give respectively 
the shallower and the steepest IMF-$\sigma$ relations.
When we fix the $T_{\rm eff, RGB}$ we still recover the same IMF--$\sigma$ relation  within the errors.  
Best fit ages, [$\alpha$/Fe] and IMF slopes obtained with $\Delta {\rm T_{\rm eff, RGB}}=-100K$ 
are consistent within $1-\sigma$ with the ones obtained by letting the $T_{\rm eff, RGB}$ be a free parameter. 
However the best-fit IMF slopes obtained with solar-scaled $T_{\rm eff, RGB}$  are systematically higher, 
especially for the two most massive bins (set 9 : $x=3.3$ for $\sigma=310\kms$ and $x=3.3$ for $\sigma=270\kms$). 
The latter result is true for both the chosen sets of indicators and is more extreme for set ID 9. 
This happens probably because by allowing the temperature of the RGB to be colder, 
the measured equivalent widths of the gravity-sensitive features will be larger, 
and therefore the inferred IMF slope will be larger.

\begin{table*}
\caption{Summary of the parameters of the inferred best-fit IMF slope as a function of central stellar velocity dispersion 
(though the linear equation $x = a\, \log\,\sigma_{200} \,+ \,b$) using different sets of indices. 
The first row represents our preferred set. The non-universality of the IMF slope is confirmed in all 
the tested combinations. However, the trend of the IMF slope with $\sigma_{\star}$ 
gets steeper when NaD is used. See the text for further details.} 
\centering 
\begin{tabular}{clrr} 
\hline
\hline 
{\bf Set ID}  &{\bf Used indices}  & {\bf a} & {\bf b}\\
{\bf (see Fig.\ref{fig:leastsquare_a}) }  &{\bf  }  & {\bf  } & {\bf  }\\
\hline 
\hline
1 & $H\beta$, Mg$b$, Fe5270,Fe5335, bTiO, aTiO,TiO1, TiO2, CaH1 and CaH2 			& $2.3\pm 0.1$  & $2.1 \pm 0.2$ \\
2 & $H\beta$, Mg$b$, Fe5270,Fe5335, bTiO, aTiO,TiO1, TiO2, CaH1 							& $2.3\pm 0.1$  & $2.1 \pm 0.2$ \\
3 & $H\beta$, Mg$b$, Fe5270,Fe5335, bTiO, aTiO,TiO1, TiO2, CaH2 							& $2.3\pm 0.2$  & $2.1 \pm 0.2$ \\
4 & $H\beta$, Mg$b$, Fe5270,Fe5335, bTiO, aTiO,TiO1, TiO2										&   $2.2 \pm 0.3$  &  $2.3 \pm 0.3$    \\
5 & $H\beta$, Mg$b$, Fe5270, bTiO, aTiO,TiO1, CaH1													&   $2.3 \pm 0.3$  &  $2.1 \pm 0.3$    \\
6 & $H\beta$, Mg$b$, Fe5270, bTiO, aTiO,TiO2, CaH2													&   $2.4 \pm 0.3$  &  $2.2 \pm 0.3$    \\
7 & $H\beta$, Mg$b$, Fe5270,Fe5335, bTiO, aTiO,TiO1, TiO2, CaH1, CaH2 and NaD 	&   $2.6 \pm 0.2$  &  $2.3 \pm 0.2$    \\
8 & $H\beta$, Mg$b$, Fe5270,Fe5335, bTiO, aTiO,TiO1, TiO2, CaH1 and NaD 			&    $2.9 \pm 0.2$  &  $2.1 \pm 0.1$     \\
9 & $H\beta$, Mg$b$, Fe5270,Fe5335, bTiO, aTiO,TiO1, TiO2, CaH2 and NaD 			&  $3.2 \pm 0.2$  &  $2.2 \pm 0.2$    \\
10 & $H\beta$, Mg$b$, Fe5270,Fe5335, bTiO, aTiO,TiO1, TiO2,  and NaD 					&   $3.2 \pm 0.2$  &  $2.4 \pm 0.2$    \\
\hline
\hline 
\end{tabular} 
\label{tab:imf_sdss_indicator}
\end{table*}

\subsection{Comparison with other works}
In this section we compare our spectroscopic stellar population-based results on the IMF--$\sigma_{\star}$ relation with 
results obtained from similar SSP-based methods by other authors and from dynamically-based and lensing-based results. 

\citet[hereafter F13]{Ferreras2013} and \citet[hereafter LB13]{LaBarbera2013} 
also found a correlation between central
velocity-dispersion and IMF slope for a large sample of SDSS
ETGs. F13 combined spectra of $\sim 40,000$ galaxies in
velocity-dispersion bins, obtaining a final set of 18 stacked spectra
at very high signal-to-noise ratio ($S/N \sim 400$ \AA\,$^{-1}$).
They compared spectral line strengths sensitive to age, metallicity,
and IMF slope of these galaxies with the population synthesis models
of Vazdekis et al.\ (2012).  Using TiO1, TiO2 and NaI ($\lambda 8190$
\AA), they found $x =4.87 \, \log\,\sigma_{200} +2.33$.
Qualitatively the general trends are similar to what we have found above, 
but their analytical fit has a much steeper slope and therefore predicts a larger variation of
the IMF slopes with galaxy mass.  For galaxies with $\sigma \geq
300\,\mathrm{km}\,\mathrm{s}^{-1}$, they inferred a very steep IMF slope of $x \sim 3.2$. 
The reason of this apparent discrepancy between our result and that of in F13 
appears to be due to the fact that F13 using SSP models with the standard 
$T_{\rm eff, RGB}$ from the isochrones. 
When using the standard $T_{\rm eff, RGB}$ we obtain an IMF-$\sigma$ 
relation that is consistent within $1\sigma$ with that of F13. 

%; this result
%violates the upper limit set by strong gravitational lensing by a very
%massive lens ETG, which ruled out IMFs with slopes steeper than
%$x=3.0$ (\citealt{Spiniello2012}).
F13 only used three indicators to constrain the stellar population parameters,
two of them coming from TiO lines that are sensitive to
$\alpha$-enrichment, as shown in our analysis.  
Disentangling IMF variations from age, metallicity and [$\alpha$/Fe] variations and
breaking the degeneracies is very difficult if only a few
IMF-sensitive features are used.  We therefore believe that our
result is more robust because it is based on the combination of more IMF-sensitive indices, 
and it infers IMF slopes that do
not violate lensing constraints on the total masses of massive ETG
lenses (see \citealt{Spiniello2011, Spiniello2012}).
In fact, LB13 find a relation similar to that we found in this work 
(within $1\sigma$, see Fig.~\ref{fig:IMF_sigma}) when analysing a variety of spectral indices, 
combining IMF-sensitive features with age- and metallicity-sensitive indices and 
considering the effect of non-solar abundance variations.
 A very good agreement is found between this work and LB13 when using a similar set of indicators (i.e. including NaD). 
LB13 and F13 test two different cases of a single power-law (unimodal) 
and a low-mass ($<0.5\,M_{\odot}$) tapered IMF (bimodal), showing that bimodal IMF shapes
do not provide such high $\Upsilon_{\star}$ ratios even for very steep slopes. 
%Here we only compare our result against their unimodal case 
%because we restrict our analysis to a single power-law IMF.

In order to compare our SSP-based results to lensing and dynamics-based results, and our unimodal IMF to the bimodal IMF of LB13,
we translate the IMF slope--$\sigma_{\star}$ relation into a ${\Upsilon}_{\star}$--$\sigma_{\star}$ relation. 
For each SDSS velocity-dispersion bin, we calculate the stellar ${\Upsilon}_{\star}$ in the $R$-band using the Dartmouth Stellar
Evolution Program (\citealt{Chaboyer2001}), selecting the IMF slope, age, and [$\alpha$/Fe] inferred from the line-strength analysis. 
We then calculate the ratio of this value and the $\Upsilon_{\star}$ that the same population (i.e. same age and [$\alpha$/Fe]) will
have assuming a Salpeter IMF (a single power law with $x=2.35$). 
In this way, we define an ``IMF mismatch" parameter $\alpha_{\rm IMF}$,
following an approach similar to the one proposed by \citet{Treu2010} and used by \citet{Cappellari2013}:

\begin{equation}
\alpha_{\rm IMF} = \dfrac{(\Upsilon)^{\star}_{x}}{(\Upsilon)^{\star}_{\rm Salp}} \, .
\end{equation}

\begin{figure*}
\includegraphics[height=9cm]{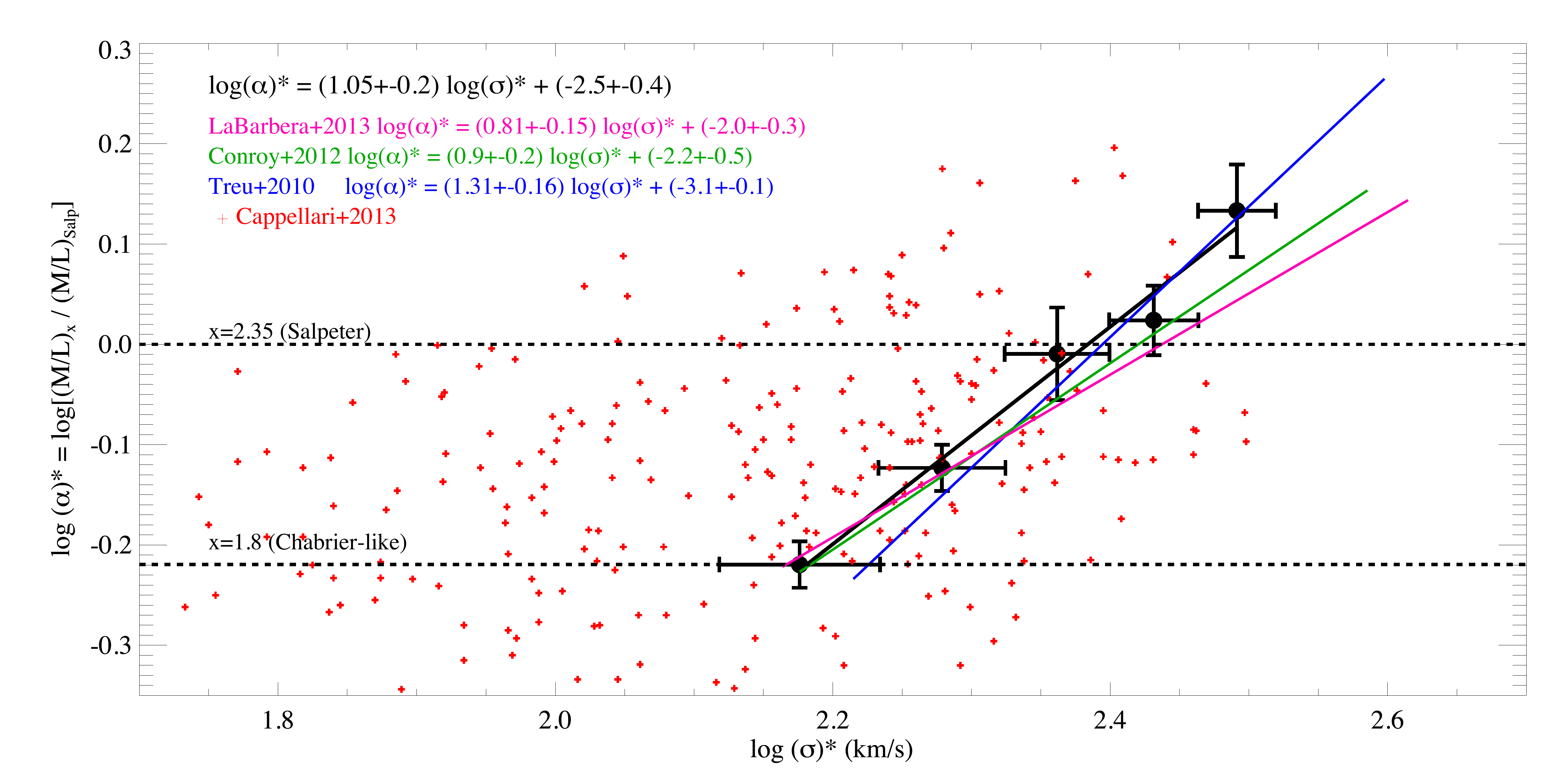}
\centering
\caption{ The $\alpha_{\rm IMF}$-$\sigma_{\star}$ relation. Our points and their error bars are shown in black, 
and a linear least-squares fit to these points is shown as the black line.
Red points are single galaxies from the ATLAS3D sample. Colored lines are least-squares fits to the 
$\alpha_{\rm IMF}$--$\sigma_{\star}$ relation derived from the spectroscopic analysis of CvD12b, of LB13 with a bimodal IMF slope, 
and the lensing+dynamic analysis from \citet{Treu2010}. 
The plot highlights the excellent agreement between %our study, LB13 \citet{Treu2010}, CvD12b.
%In general, we find good agreement between 
these independent studies in their common velocity dispersion range. 
The Chabrier and Salpeter cases are shown as horizontal dotted lines. 
A bottom-heavy IMF with slope of $x=3.0$ predicts  $\log(\alpha_{\rm IMF}) \sim 0.6$.}
\label{fig:MoL_comparison}
\end{figure*}

In Figure~\ref{fig:MoL_comparison} we plot $\log (\alpha_{\rm IMF})$ 
as a function of $\log \sigma_{\star}$ and compute the corresponding linear 
regression for the SDSS points (black points)
\begin{equation}
\log (\alpha)_{\rm IMF} = (1.05\pm 0.2) \, \log\,\sigma_{\star} - (2.5\pm 0.4) \, .
\end{equation}
\noindent Moreover, we show linear fits to the gravitational-lensing-based results of SLACS (blue, \citealt{Treu2010}) 
and to stellar population-based results (green: \citealt{Conroy2012b}; 
magenta: \citealt{LaBarbera2013}). Finally we plot the sample of 260 ATLAS3D galaxies 
(red points, \citealt{Cappellari2011, Cappellari2013}). 
The linear fit of LB13 is obtained using the linear relation between best-fit slope of the IMF 
and central velocity dispersion given in their Figure~12. 
We use the mass-to-light ratios in the Johnson $R$-band (Vega system) predicted from the MIUSCAT SSP
models\footnote{http://miles.iac.es/pages/photometric-predictions/based-on-miuscat-seds.php} with an age of 12.6 Gyr, 
solar metallicity and bimodal IMFs to convert the IMF slope--$\sigma_{\star}$ relation into 
a ${\Upsilon}_{\star}$--$\sigma_{\star}$ relation. 
In particular we use the bimodal IMF slopes obtained by fitting spectral indices with two SSP plus X/Fe models. 
These models consist of a linear combination of two extended MIUSCAT SSPs with the same
IMF but different ages and metallicities. In addition, the models incorporate 
three free parameters describing the [Ca/Fe], [Na/Fe], and [Ti/Fe] abundances. 

In Figure~\ref{fig:MoL_comparison}, we also draw two horizontal lines showing the $\alpha_{\rm IMF}$ for a Chabrier-like 
(single power-law with $x=1.8$) and a Salpeter IMF (i.e. $\log (\alpha_{\rm IMF})=0$).
We note that a bottom-heavy IMF with slope of $x=3.0$ falls out of the plot, predicting a value 
of $\log (\alpha_{\rm IMF}) \sim 0.6$.
The good agreement between the stacked SDSS data points and these completely independent 
analyses (at least at the massive end of the different samples) gives confidence that our spectroscopic 
analysis and our inferred variation of the IMF slope with velocity dispersion
are robust, and the trend we observe depends neither on the SSP model chosen nor on our statistical analysis.

%\section{Discussion}
%As already pointed out by \citet{Conroy2012b} (see their nice Fig.12), 
%Qualitatively, we recover the same trends no matter which
%set of IMF-sensitive features is used. In particular, in all cases
%we find evidence for IMF variation. This provides strong
%confirmation that our basic result is not influenced by a single
%spectroscopic feature. There are however noticeable offsets
%between the various permutations, especially in the upper left
%panel of Figure 12. This means that in detail the various
%IMF-sensitive features are favoring different values of M/LK,
%which in turn suggests that there are residual systematics in the
%modeling. One possible explanation lies in the fact that these
%features are each most sensitive to a different range of stellar
%masses (see CvD12 for details). The data may be demanding an
%IMF that is more complicated than the broken power law that
%we adopt. These issues will be explored in future work.
%

\section{Discussion and Conclusions}

In this paper, we have defined a new set of indices which are strong in cool giants and dwarfs and
almost absent in main sequence stars, in a spectral region where single-stellar population (SSP) models have
been most extensively studied.
These features arise from TiO and CaH molecular absorption bands in
the wavelength range $\sim 4700$--$7000$ \AA.  We have calculated the
strengths of these indices for stars in the MILES empirical library.
In particular we have shown that all the TiO indicators are strong in
cool giants and dwarfs and almost absent in warm main-sequence
  stars, allowing the study of their variation with temperature,
gravity and [Fe/H].  However, the strength of the TiO lines increases with the increasing 
of the IMF but also with increasing [$\alpha$/Fe]. 
The feature around CaH1 $\lambda$6380, which is strong only in M dwarfs, 
decreases instead with increasing $\alpha$-enhancement and therefore allows us,
in combination with the TiO indicators, to break the degeneracy between IMF 
and [$\alpha$/Fe] variations and 
thus to constrain the low-mass end of the IMF slope. 

We have used the \citet{Conroy2012} SSP models constructed specifically for the purpose of
measuring the IMF slope down to $\sim0.1\,M_{\odot}$ for old,
metal-rich stellar populations.  We have measured SSP index strengths
for our new set of IMF indicators, and we have compared these to
strengths of stacked spectra of SDSS DR8 ETGs in five different velocity
dispersion bins between $150$ and $310\,\mathrm{km}\,\mathrm{s}^{-1}$.
Finally we also investigated in detail the response of a change in the
effective temperature of the red giant branch in the isochrones versus
the response in variation of the IMF slope for the wavelength range
where these responses are shown to be mostly degenerate.

In this work, we focused only on line-index measurement and we restricted
the analysis to solar metallicity.  Element abundance
is a fundamental parameter that has to be fully and quantitatively
explored in order to further isolate and test the suggested
variation of IMF normalization with galaxy mass.  A full spectral
fitting approach can help to further investigate possible IMF
variations with galaxy masses and to trace the complete star formation
history of a galaxy (\citealt{Conroy2013b}).

Our main conclusions are the following.
\begin{itemize}
\item CvD12 SSP models predict a minimal variation of the aTiO and CaH1 EWs 
with age (at least  for old ages) and a minimal dependence of
  [$\alpha$/Fe] on TiO2.
\item The predicted variation of IMF slopes is orthogonal to 
  [$\alpha$/Fe] enrichment for all the presented IMF-sensitive
  index-index plots.
\item Our set of IMF-sensitive indicators is able to break the
  IMF--${T_{\rm eff, RGB}}$ degeneracy.  The index strengths of SDSS
  galaxies match better those of SSP models with a population of RGB
  stars slightly cooler than the default isochrones.  However this cannot
  explain the trend of IMF slope with galaxy velocity dispersion.
\item A variation of the IMF with galaxy velocity dispersion,
  consistent with previous works (\citealt{Treu2010, Spiniello2011, Spiniello2012, Cappellari2012, Tortora2013,LaBarbera2013})
 is visible in all the index-index plots for the SDSS stacked spectra of galaxies in
  different $\sigma_{\star}$ bins from 150 to $310\,\mathrm{km}\,\mathrm{s}^{-1}$.  
  We find $x = (2.3\pm 0.1) \, \log\,\sigma_{200} + (2.13\pm 0.15)$ using ten indicators.
  %in disagreement with \citet{Ferreras2013}, who found a significantly steeper relation when using an IMF characterized by a  single power-law, but 
 This result is within $1\sigma$ of \citet{LaBarbera2013}, who performed a detailed 
  analysis using more IMF-sensitive indicators than \citet{Ferreras2013}, two 
SSP models and a correction for the effect of non-solar abundance patterns (X/Fe). 
\item Our fit predicts an IMF slope that is more shallow than 3.0 for the most massive 
  $\sigma$-bin.  This is in agreement with the upper limit sets by 
  gravitational lensing studies on the analysis of one very massive lens ETG (\citealt{Spiniello2012}) 
\item We translate the IMF--$\sigma_{star}$ relation into a ${\Upsilon}s_{\star}$--$\sigma_{\star}$ 
relation and compute an ``IMF mismatch" parameter ($\alpha_{\rm IMF}$). 
This allows us to compare our results with \citet{Cappellari2013}, \citet{Treu2010} \citet{LaBarbera2013} and \citet{Conroy2012b}.  
The studies, based on three completely independent methods, and two different SSP models are in excellent agreement. 
\end{itemize}

In conclusion, the newly defined optical IMF indicators bTiO, aTiO, TiO1, TiO2, CaH1, and CaH2, 
represent useful probes for decoupling the IMF from stellar population age,
metallicity, bulk abundance ratio [$\alpha$/Fe], and the effective temperature of the RGB, when used in
combination with age and metallicity indicators.
These features are in a region of the spectrum less affected by sky
lines or atmospheric absorption lines than the NIR features previously
used.  This region is also easier to observe as it lies in the
wavelength region of the majority of the optical spectrographs,
including integral-field instruments, that could be used to spatially
resolve the low-mass stellar population of galaxies (Spiniello et al. 2014, in prep.).

In this work, we focussed only on line-index measurements and we restricted our analysis 
to solar metallicity and abundances. 
In this context, results relying primarily on NaD should be treated with some caution.
Element abundance is a fundamental parameter that has to be fully and quantitatively 
explored in order to completely isolate and test the suggested variation of IMF 
normalization with galaxy mass (in particular [Na/Fe], Spiniello et al., 2013b, in prep.).
A full spectral fitting approach might also be necessary to further investigate possible IMF variations with galaxy 
masses and to trace the complete star formation history of a galaxy (\citealt{Conroy2013b}).

\section*{Acknowledgements}
C.S. acknowledges support from an Ubbo
Emmius Fellowship. C.C. acknowledges support from the Alfred P. Sloan Foundation. 
C.S. thanks Amina Helmi, Gerjon Mensinga, Matteo Barnab\`e and Gergo Popping for thoughtful comments 
that have improved the quality of the manuscript.
We thank Michele Cappellari, Sukyuong Ken Yi and Hyunjin Jeong for interesting discussions and suggestions.
%The authors thank Charlie Conroy for assistance and help with using and interpreting his models.
We are grateful to the anonymous referee for his/her insightful and constructive comments.
Funding for SDSS-III has been provided by the Alfred P. Sloan Foundation, the Participating Institutions, 
the National Science Foundation, and the U.S. Department of Energy Office of Science. 
The SDSS-III web site is http://www.sdss3.org/.
SDSS-III is managed by the Astrophysical Research Consortium for the Participating Institutions of the SDSS-III 
Collaboration including the University of Arizona, the Brazilian Participation Group, 
Brookhaven National Laboratory, University of Cambridge, Carnegie Mellon University, University of Florida, the 
French Participation Group, the German Participation Group, Harvard University, the Instituto de Astrofisica de Canarias, 
the Michigan State/Notre Dame/JINA Participation Group, Johns Hopkins University, Lawrence Berkeley National Laboratory, 
Max Planck Institute for Astrophysics, Max Planck Institute for Extraterrestrial Physics, New Mexico State University, New York 
University, Ohio State University, Pennsylvania State University, University of Portsmouth, Princeton University, the Spanish 
Participation Group, University of Tokyo, University of Utah, Vanderbilt University, University of Virginia, University of Washington, and Yale University.

%%%%%%%%%%%%%%%%%%%%%%%%%%%%%%%%%%%%%%%%%%%%%%%%%%%%%%%%
% INPUT BIBLIOGRAPHY
%%%%%%%%%%%%%%%%%%%%%%%%%%%%%%%%%%%%%%%%%%%%%%%%%%%%%%%%
\bibliographystyle{../mn2e_fix}
\bibliography{Spiniello13_REVISED_Astroph}

\label{lastpage}

\end{document}